\documentclass[useAMS,usenatbib,longnamesfirst,usegraphicx]{mn2e}

\usepackage{amssymb}

\shortcites{Astier2006,Baron1996a,Baron2006,Blinnikov1998,Blinnikov2006,Branch1982,Branch1983,Gamezo2003,Hoeflich1996,Jeffery1992,Karp1977,Krisciunas2003,Lentz2001,Nugent1997,Patat1996,Perlmutter1999,Roepke2007a,Riess1998,Riess2004,Salvo2001,Sim2007a,Verner1996,Wood-Vasey2007,Sauer2008a}

\title[Time-dependent 3D spectrum synthesis for type Ia supernovae]{
Time-dependent 3D spectrum synthesis for type Ia supernovae}
\author[M. Kromer and S. A. Sim]{
M. Kromer\thanks{E-mail: mkromer@mpa-garching.mpg.de} and S. A. Sim\\
Max-Planck-Institut f\"ur Astrophysik, Karl-Schwarzschild-Stra{\ss}e 1, D-85748 Garching b. M\"unchen, Germany}
\begin{document}

\date{16. June 2009}

\pagerange{\pageref{firstpage}--\pageref{lastpage}} \pubyear{}

\maketitle

\label{firstpage}

\begin{abstract}
  A Monte Carlo code (\textsc{artis}) for modelling time-dependent three-dimensional 
  spectral synthesis in chemically inhomogeneous models of Type Ia 
  supernova ejecta is presented. Following the propagation of $\gamma$-ray 
  photons, emitted by the radioactive decay of the nucleosynthesis products,
  energy is deposited in the supernova ejecta and the radiative transfer
  problem is solved self-consistently, enforcing the constraint of energy 
  conservation in the co-moving frame. Assuming a photoionisation dominated
  plasma, the equations of ionisation equilibrium are solved together 
  with the thermal balance equation adopting an approximate treatment of
  excitation. Since we implement a fully general treatment of line formation, 
  there are no free parameters to adjust. Thus a direct comparison between 
  synthetic spectra and light curves, calculated from hydrodynamic explosion 
  models, and observations is feasible. The code is applied to the well known
  W7 explosion model and the results tested against other studies. Finally 
  the effect of asymmetric ejecta on broad band light curves and spectra is 
  illustrated using an elliptical toy model.
\end{abstract}


\section{Introduction} 
Type Ia supernovae (SNe Ia) are commonly believed to originate from 
thermonuclear explosions of degenerate material in white dwarf stars 
\citep{Hoyle1960}, thus being important for binary evolution and
the chemical evolution of the Universe. By their dynamical interaction
with the interstellar medium they may also play an important role in the
star formation history of galaxies. However, even after many years of 
intensive research it is still not completely understood how these 
explosions take place and which are the progenitor systems [see e.g.
\citet{Hillebrandt2000} for a review on the explosion mechanism and 
\citet{Livio2000} on progenitors].

In the standard single degenerate Chandrasekhar mass model, a C+O white dwarf 
accretes mass from a non-degenerate companion star (either a main-sequence or 
red giant star) until it nears the Chandrasekhar mass. Then the pressure of 
the degenerate electrons can no longer balance gravity and a thermonuclear 
runaway, disrupting the white dwarf, can occur. An attractive feature of this
picture is that exploding at a well specified mass provides a natural 
explanation of the observed homogeneity of the SNe Ia class.

However, SNe Ia are not all the same. Observations of nearby SNe Ia show 
a broad scatter in peak luminosity and decay time-scale. \citet{Phillips1993}
derived empirical relations to calibrate the peak luminosity of SNe Ia by 
distance independent light curve properties which makes them standardiseable 
candles. Because of their high luminosities (typically 
$\sim10^{42}\,\mathrm{erg\,s}^{-1}$) they can thus  be used to measure the 
cosmic expansion history up to high redshifts. Various forms of this approach 
have been used by different groups \citep[e.g.][]{Riess1998,Perlmutter1999}. 
They found that SNe Ia at high redshift appear fainter than expected in a matter 
dominated Universe, indicating an accelerated expansion. This was confirmed 
by following studies \citep[e.g.][]{Riess2004,Astier2006} and led to the 
$\Lambda$CDM cosmologies in which a positive cosmological constant or 
``dark energy'' with negative pressure is used to model the accelerated 
expansion of the Universe.

One way to constrain dark energy models is to study the evolution of the dark
energy equation-of-state parameter $w$ by measuring the cosmic expansion history,
a goal of ongoing [e.g. ESSENCE \citep{Wood-Vasey2007}, SNLS \citep{Astier2006}] 
and future projects. For that purpose, the distances to high-redshift SNe Ia 
must be determined to high accuracy, which requires, amongst other factors 
(e.g. the extinction to the supernova and K-corrections), a reliable calibration 
of the peak luminosity. To improve the calibration techniques, which are based 
on purely empirical relationships, a thorough theoretical understanding of SNe 
Ia and improved observations are needed.

Much progress (both in numerical methods and computational power) has been 
made in the hydrodynamical explosion modelling of SNe Ia since the first 
one-dimensional (1D) calculations by e.g. \citet{Nomoto1984} became 
available. Today, fully three-dimensional (3D) explosion models  
\citep[e.g.][]{Reinecke2002,Gamezo2003,Roepke2005a,Roepke2007} are the 
state-of-the-art and have shown that 3D effects are essential to properly 
simulate the instabilities and turbulence effects which drive the thermonuclear 
combustion. Furthermore they show that ejecta asymmetries can arise, either 
by hydrodynamical instabilities during the burning phase or an asymmetric 
ignition \citep{Hoeflich2002,Kuhlen2006}.

However these simulations -- which give velocities, densities and composition 
of the explosion ejecta -- are not directly comparable to observations of real 
SNe Ia. For that purpose synthetic spectra and light curves must be obtained 
by radiative transfer calculations. This requires a solution of the multi-line
transfer problem in expanding media where the opacity is dominated
by the wealth of lines associated with the iron group elements which
were synthesised in the thermonuclear explosion. Many 1D studies have 
addressed this problem in the past either assuming pure resonance scattering 
\citep[e.g.][]{Branch1982,Branch1983,Mazzali1993} 
or pure absorption \citep[e.g.][]{Jeffery1992} in the lines. 
\citet{Lucy1999b} introduced an approximate treatment of line fluorescence. 
But that work is still in 1D as are the studies done with the general-purpose 
radiation transport code \textsc{phoenix} \citep[e.g.][]{Lentz2001}. Recently following 
\citet{Lucy2005}, \citet{Kasen2006} described a 3D time-dependent radiative 
transfer code which is capable of treating line fluorescence in an approximate 
way \citep[similar to][]{Lucy1999b,Pinto2000b}.

Such studies have shown that to address the complexity of the hydrodynamic 
explosion models a 3D treatment of radiative transfer which simulates the 
$\gamma$-deposition and spectrum formation in detail is needed. In particular, 
a careful treatment of the ionisation balance and a proper simulation of the 
redistribution of flux by line fluorescence [crucial for the near-infrared 
light curves, see \citet{Kasen2006a}] is essential. Here we present a new 
Monte Carlo code (\textsc{artis}, Applied Radiative Transfer In Supernovae) which 
(based on the approach of \citealt{Lucy2002,Lucy2003,Lucy2005}) solves the 
time-dependent 3D radiative transfer problem in chemically 
inhomogeneous models of supernova ejecta from first principles using a 
generalised treatment of line formation and prioritising a detailed treatment 
of ionisation. The radiative transfer calculation is parameter-free, depending 
only on the input model and atomic data, giving a maximum of predictive power 
for a given hydro model. Details of our code are given in Section \ref{sec:method}.

In Section \ref{sec:w7} we use this code to calculate model spectra and 
light curves for the well-studied 1D deflagration model W7 \citep{Nomoto1984}
and compare them to observations and earlier synthetic results in order to test 
our code. We also investigate the influence of completeness of atomic data. Finally,
in Section \ref{sec:elmodel}, line-of-sight dependent spectra and light curves 
for an ellipsoidal toy model are calculated to demonstrate the multi-dimensional
capabilities of our code and study the effect of large scale asymmetries 
of the ejecta, before we draw conclusions in Section \ref{sec:conclusions}.

\section{Method} 
\label{sec:method}
Monte Carlo methods have been widely applied to model astrophysical radiative 
transfer in expanding media (e.g. \citealt{Abbott1985} and \citealt{Lucy1993} 
for winds of massive stars and \citealt{Mazzali1993} for the expanding envelopes 
of supernovae). This is mainly due to the ease with which they address the 
multi-line transfer problem -- a single photon can interact with many spectral 
lines due to the progressive redshift it experiences in the co-moving frame 
while it travels through the expanding envelope. Monte Carlo methods are also
readily extended to multi-dimensional problems (see e.g. \citealt{Kasen2006} 
and \citealt{Maeda2006}) and easy to parallelise, allowing the exploitation 
of massively parallel super computers.

As the typical expansion velocities in supernova envelopes ($\lesssim 
30\,000\,\mathrm{km\,s}^{-1}$) imply large velocity gradients, the physical 
properties of the envelope will not change significantly within the resonance 
region of a spectral line (Doppler velocity $v_\mathrm{D}\sim5\,\mathrm{km\,s}^{-1}$) 
and we adopt the Sobolev approximation and consider that line absorption of 
photons takes place only at a particular point of resonance. This simplifies 
the problem significantly and is a fundamental assumption of our study. 
For a detailed description of the Sobolev approximation see e.g. 
\citet{Mihalas1978} or \citet{Lamers1999}.

The second fundamental assumption we make is that of homologous expansion, 
thereby decoupling the radiative transfer from the hydrodynamic evolution
of the ejecta. This applies if the ejecta are in free expansion, i.e.
their kinetic energy density dominates the gravitational and internal 
energy densities. For SNe Ia this is achieved less than a minute after the 
explosion \citep[e.g.][]{Roepke2005}. Thus it should be an excellent approximation 
for our radiative transfer calculations which are usually started at $\sim2$ 
days. In homologous expansion the velocity $v$ of the ejecta at a particular
position $r$ is always proportional to the position, $v=r/t$ with $t$ being
the time since explosion.

\subsection{Monte Carlo radiative transfer}
We extended the 3D Monte Carlo radiative transfer code introduced by 
\citet{Sim2007} to a non-grey opacity treatment following the scheme 
outlined in a series of papers by \citet{Lucy2002,Lucy2003,Lucy2005}. 
This scheme is based on the artificial division of the radiation field
into indivisible energy packets as Monte Carlo quanta, rather than 
Nature's quantisation of radiation. This has two major advantages: first 
it keeps the code simple as packet histories can be followed one-by-one, 
avoiding the need to follow multiple packets in e.g. recombination cascades.
Second it ensures rapid convergence to an accurate temperature stratification 
by imposing the constraint of energy conservation in the co-moving frame 
\citep{Lucy1999a}.

\begin{figure*}
  \centering
  \includegraphics{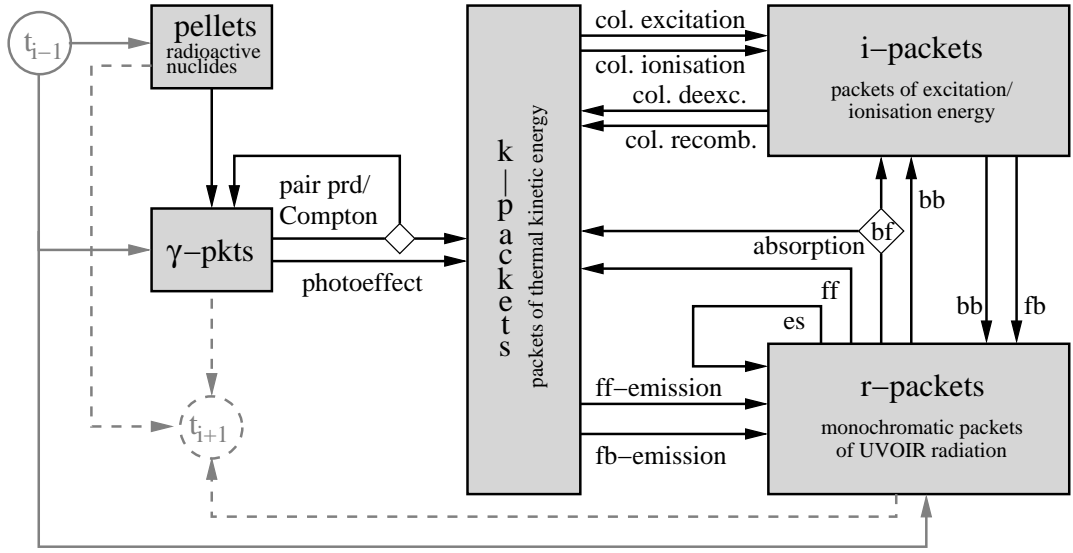}
  \caption{Flow chart outlining the mode of operation of the code. For discussion see text.}
  \label{fig:flowchart}
\end{figure*}

In the following we outline the operation of the code with reference to the
flow chart shown in Figure \ref{fig:flowchart} and discuss the relevant physical 
processes, giving detail only where we deviate from the formulations used by 
\citet{Lucy2002,Lucy2003,Lucy2005}. The basic transfer calculations are carried 
out in the rest frame (rf, quantities always denoted unprimed) so that an energy 
packet's trajectory at time $t$ is described by its position $\bmath{r}(t)$ and 
direction $\bmath{\mu}(t)$. For interactions with matter we transform to the local 
co-moving frame (cmf, quantities denoted with prime). 

\subsubsection{Setting up the computational domain}
To avoid any symmetry assumptions and keep the code simple, we map the explosion
ejecta to a 3D Cartesian grid with origin at the centre-of-mass of the supernova
and containing $I^3$ cubic cells. The grid expands with time to follow the 
evolution of the ejecta. Physical variables are assumed not to vary spatially 
within the grid cells. As input we take densities $\rho_i(t)$, velocities 
$v_i(t)$ and composition $X_{i,k}(t)$ from explosion models specified for
the phase of homologous expansion and map these for a time $t_0$ onto the 
grid. $t_0$ and all time specifications in the following indicate the time 
since explosion, unless otherwise noted.

We then follow the expansion of the ejecta for $N$ time steps up to time $t_N$ 
by expanding the individual grid cells continuously. The time steps 
($t_n$,$t_{n+1}$) are spaced logarithmically and thermodynamic quantities in 
a cell such as densities, temperatures and atom/ion populations are kept fixed 
during a time step $n$ at the value they have for $t_{n+0.5}$.

\subsubsection{Energy deposition}
SNe Ia light curves are primarily powered by the radioactive decays of
$^{56}$Ni$\rightarrow^{56}$Co and $^{56}$Co$\rightarrow^{56}$Fe 
\citep{Truran1967,Colgate1969} giving rise to the emission of a spectrum 
of $\gamma$-photons associated with their transitions \citep{Ambwani1988}. 
Out of the total energies, $E_\mathrm{Ni}$ and $E_\mathrm{Co}$ emitted 
per decay of $^{56}$Ni and $^{56}$Co, we determine the total $\gamma$-ray 
energy emitted in the decay chain for $t\rightarrow\infty$
  \begin{equation}
    E_\mathrm{tot}=(E_\mathrm{Ni}+E_\mathrm{Co})M_\mathrm{Ni}/m_\mathrm{Ni}.
  \end{equation}
$M_\mathrm{Ni}$ is the initial mass of $^{56}$Ni synthesised in the explosion 
and $m_\mathrm{Ni}$ the mass of the $^{56}$Ni atom.

Following \citet{Lucy2005}, this energy is quantised into 
$\mathcal{N}=E_\mathrm{tot}/\epsilon_0$ identical energy packets of cmf 
energy $\epsilon_0$ which we call ``pellets'' in the following. These pellets 
are distributed on the grid according to the initial $^{56}$Ni distribution 
and follow the homologous expansion until they decay. Decay times are sampled
randomly according to the $^{56}$Ni$\rightarrow^{56}$Co$\rightarrow^{56}$Fe 
decay chain with pellets assigned to represent either the $^{56}$Ni or 
$^{56}$Co decay.

Upon decay, a pellet transforms to a single $\gamma$-packet representing a 
bundle of monochromatic $\gamma$-radiation with cmf energy $\epsilon_0$
and a cmf photon energy $E'_\gamma$ which is randomly sampled from the 
$\gamma$-lines in the appropriate decay of the sequence 
$^{56}$Ni$\rightarrow^{56}$Co $\rightarrow^{56}$Fe. The $\gamma$-packet's 
direction ($\bmath{\mu'}$) in the cmf is sampled randomly assuming 
isotropic emission.

\subsubsection{Propagation of $\gamma$-packets}
The $\gamma$-packets are propagated through the ejecta in the rest frame of 
the grid until either (i) they leave the grid, (ii) the current time step 
finishes or (iii) they interact with matter. Computing the distances to all 
possible events along the packet trajectory, we select the event which is 
reached first [see \citet{Lucy2005} for details]. In the first case the 
$\gamma$-packet is flagged inactive and the calculation proceeds to the 
next active packet. In the second case we save the rf data string
($\bmath{r}$, $t$, $\bmath{\mu}$, $\epsilon$, $E_\gamma$) for the following 
time step and continue with the next active packet.

For the third case we consider Compton scattering by free and bound 
electrons, photoelectric absorption and pair production as possible physical 
processes, the last being only available for $\gamma$-packets with energy
$E>2m_\mathrm{e}c^2$. For details see again \citet{Lucy2005} and Equation 1 
of \citet{Sim2008} for the adopted photoabsorption cross-section. Which of the 
processes happens is determined randomly according to their absorption 
and scattering coefficients. 

In the case of a photoelectric absorption, the $\gamma$-packet energy is 
deposited as thermal kinetic energy. In the framework of this code, this 
is described by a transformation of the $\gamma$-packet into a so-called 
$k$-packet of equal cmf energy. The treatment of $k$-packets is described
in section \ref{sec:kpackets}.

As our energy packets are indivisible, the treatment of Compton scattering
and pair production, where the photon energy is distributed to two particles,
is slightly more complex. Following again \citet{Lucy2005}, for Compton 
scattering the $\gamma$-packet is either scattered and continues as a 
$\gamma$-packet of the same cmf energy as the incident packet or it is 
transformed into a non-thermal $e^-$-packet. $e^-$-packets are assumed 
to thermalise and are instantaneously transformed into $k$-packets.
For pair production we either create $e^+$- or $e^-$-packets. Assuming 
in situ annihilation, for an incident $\gamma$-packet of cmf photon energy 
$E'_\gamma$ a fraction of $2m_\mathrm{e}c^2/E'_\gamma$ (represented by 
the $e^+$-packets) is released in form of $\gamma$-rays at 0.511\,MeV 
when the positron annihilates. The remainder (representing the kinetic 
energy of the electrons and positrons) goes directly to the thermal pool.

\subsubsection{Treatment of thermal kinetic energy}
\label{sec:kpackets}
Neglecting energy storage in the ejecta gas, thermal kinetic energy converts
instantaneously (i.e. without propagating) into ultraviolet-optical-infrared
(\textit{UVOIR}) radiation. This happens either directly via continuum 
emission by free-free or free-bound processes or indirectly by collisional
excitations/ionisations of the gas and subsequent radiative deexcitations/recombinations. 
In our framework, this means either transforming a $k$-packet into an 
$r$-packet -- representing a monochromatic energy packet of \textit{UVOIR} 
radiation (for details see Section \ref{sec:rpackets}) -- or into a packet 
of atomic internal energy, which we call $i$-packets. Details of the latter 
will be discussed in the next section.

Which of the processes happens depends on the cooling rate associated with 
the particular process and is sampled randomly. For the free-bound cooling
rate we consider only spontaneous recombination since stimulated 
recombination is treated as negative photoionisation and, as such, is taken 
into account in the creation of $k$-packets.

From sampling the cooling rates, we know in which free-bound continuum 
(specified by level $i$ of ion $j$ of element $k$) to emit or whether 
free-free emission was selected. In both cases the $r$-packet's direction 
in the cmf $\bmath{\mu'}$ is determined according to isotropic emission 
and the cmf energy $\epsilon'$ of the generating packet is conserved. After 
transforming these quantities into the rf together with the packet's 
position ($\bmath{r}$), we must determine the $r$-packet photon 
frequency before continuing its propagation through the grid.

For free-free emission the frequency is determined from
  \begin{equation}
    \int_\nu^\infty j^\mathrm{ff}(\nu')\,\mathrm{d}\nu'
    = z \int_0^\infty j^\mathrm{ff}(\nu')\,\mathrm{d}\nu'
  \end{equation}
with $z$ a random number in $]0,1[$ and $j^\mathrm{ff}(\nu)\propto\exp^{-h\nu/kT_\mathrm{e}}$ 
(we neglect the frequency dependence of the velocity-averaged Gaunt factor). 
For free-bound emission we use
  \begin{equation}
    \label{eq:fb_emissivity}
    \int_\nu^\infty j_{i,j,k}^\mathrm{fb}(\nu')\,\mathrm{d}\nu'
    = z \int_{\nu_{i,j,k}}^\infty j_{i,j,k}^\mathrm{fb}(\nu')\,\mathrm{d}\nu' 
  \end{equation}
with $z$ a random number in $]0,1]$ and where $\nu_{i,j,k}$ is the edge 
frequency of the continuum. The emissivity $j_{i,j,k}^\mathrm{fb}(\nu)$
for spontaneous recombination by that continuum is given by Equation 23
of \citet{Lucy2003}.

In principle we want to only release thermal energy when transforming a 
$k$-packet to an $r$-packet. However in emitting a bound-free photon at 
frequency $\nu$, a fraction $\nu_{i,j,k}/\nu$ of the packet energy is drawn
from the ionisation/excitation energy pool. The machinery described in the 
next section ensures that this is balanced by the amount of thermal energy 
which is released by recombination processes in our treatment of atomic 
internal energy (see also \citealt{Lucy2003}).

\subsubsection{Treatment of atomic internal energy}
\label{sec:ma}
We treat atomic internal energy using the macro-atom approach of 
\citet{Lucy2002,Lucy2003} which is based on an interpretation of 
statistical equilibrium in terms of macroscopic energy flow rates. For
Monte Carlo methods this is an improvement over earlier work in 1D 
\citep[e.g.][]{Mazzali1993,Lucy1999b} or in 3D \citep{Kasen2006}.
We stress that the macro-atom approach allows for a fully general
treatment of radiation/matter interactions in statistical equilibrium, 
including true absorption, scattering and fluorescence.

Whenever a packet is converted to atomic internal energy ($i$-packet), 
we assign it to a macro-atom state (see \citealt{Lucy2002} for relevant 
definitions). The ejecta gas can be excited/ionised either collisionally, 
represented by a $k$-packet to $i$-packet transition (see last section)
or radiatively by \textit{UVOIR} radiation, represented by an $r$-packet 
to $i$-packet transition. Non-thermal excitations by fast electrons and 
$\gamma$-rays are not currently taken into account but the method could 
be extended to include them in due course. 

For $k$-packet to $i$-packet transitions, the macro-atom state is determined
by sampling the cooling rates for collisional excitation/ionisation, which 
are calculated in the van Regemorter and Seaton approximations \citep{Mihalas1978}.
For $r$-packet to $i$-packet transitions, the absorption process identifies
the appropriate macro-atom state directly. Details of the treatment of 
$r$-packets are given in Section \ref{sec:rpackets}.

We do not take into account energy storage in the ejecta gas but assume 
$i$-packets convert instantaneously back into $k$- or $r$-packets. For that 
purpose we calculate rates for all transitions which connect an activated
state to other macro-atom states or which allow the macro-atom to
deactivate, i.e. transform the internal atomic energy back to either 
thermal kinetic ($k$-packet) or radiative ($r$-packet) energy. From all 
these transitions we randomly select one by sampling the probabilities
which are computed following \citet{Lucy2003}. If it is a macro-atom 
internal jump we reset the $i$-packet macro-atom state and continue 
this procedure until we select a deactivating process.

During deactivations we conserve the cmf energy of the packet to enforce
radiative equilibrium. In the case of a collisional deactivation the 
$i$-packet is just transformed into a $k$-packet. In case of radiative
deexcitation/recombination we transform the $i$-packet into an $r$-packet 
and emit it isotropically in the cmf. Finally we need to assign a photon 
frequency to the $r$-packet. For a radiative deexcitation this is given 
by the frequency of the associated line transition in the cmf. In the 
case of a radiative recombination, we again sample the emissivity for the 
free-bound emission (Equation \ref{eq:fb_emissivity}). Note that we thereby 
emit both ionisation and thermal energy, as we do it in the case of 
free-bound cooling.

\subsubsection{Propagation of UVOIR radiation}
\label{sec:rpackets}
$r$-packets representing monochromatic energy packets of \textit{UVOIR} 
radiation are propagated through the ejecta from their first emission by any
of the above processes until (i) they leave the grid, (ii) the current time 
step finishes or (iii) they interact with matter. In the zeroth time step 
we also have radiation which comes from pellets which decayed before $t_0$. 
We assume that the ejecta are optically thick at these early times such that 
the radiation is trapped and local thermodynamic equilibrium (LTE) is a 
good approximation. Therefore we emit these packets isotropically at the 
position the pellets would have at $t_0$ and assign them frequencies according 
to a black-body distribution at the local kinetic temperature $T_\mathrm{e}$.

We follow the packets along their trajectory until they are stopped 
by the first event which occurs. For (i) and (ii) this is a simple 
geometrical problem. For case (iii) it requires calculating the distance 
to a randomly sampled optical depth $\tau_\mathrm{r} = -\ln z$ with  
$z\in]0,1]$, as described by \citet{Mazzali1993}. First 
the trajectory point at which the photon comes into resonance with the 
next spectral line of the line list is calculated in the Sobolev 
approximation. 
If the continuum optical depth accumulated up to that point is bigger 
than the random optical depth, the travelled distance $\mathrm{d}s$ is 
calculated from the continuum opacity ($\kappa_\mathrm{cont}$) from
$\kappa_\mathrm{cont}\mathrm{d}s = \tau_\mathrm{r}$ and a continuum absorption
occurs. If the sum of continuum and line optical depth is bigger than the 
random optical depth, $\mathrm{d}s$ is given by the distance to the resonance 
point and a line event occurs. Otherwise we calculate the distance to the 
next spectral line with which the photon comes into resonance and repeat 
the procedure.

According to the chosen event, the packet properties are now updated. For 
case (i), the $r$-packet is flagged inactive and we proceed to the next 
active packet. For case (ii), we save the actual rf data string ($\bmath{r}$, $t$, 
$\bmath{\mu}$, $\epsilon$, $\nu$) for the following time step and continue 
with the next active packet.
For case (iii), we change the packet type according to the process which
terminated the packet flight. For a line absorption, we convert the 
$r$-packet into an $i$-packet and set it to the corresponding macro-atom
target state and the macro-atom machinery (Section \ref{sec:ma}) determines
its fate.

For a continuum event, we first decide whether it was Thomson scattering,
free-free absorption or bound-free absorption by randomly sampling their 
absorption coefficients. Thomson scattering, which we assume to be coherent
and isotropic, changes neither the packet type, nor the cmf frequency of the 
packet. Thus, Thomson scattering amounts to isotropic reemission of the 
$r$-packet in the local cmf. Free-free absorption transforms the absorbed 
radiative energy completely into thermal kinetic energy, i.e. converts
the $r$-packet into a $k$-packet.

Bound-free absorption is more complicated since it contributes both to 
atomic internal energy and thermal kinetic energy. After randomly selecting 
one specific continuum according to its absorption coefficient, we follow
Lucy in forcing a conversion to either a single $k$- or $i$-packet.
For an $r$-packet absorbed with frequency $\nu$ by a continuum with edge 
frequency $\nu_{i,j,k}$, a fraction $\nu_{i,j,k}/\nu$ of its energy is
converted to atomic internal energy and the rest to thermal kinetic energy. 
To account for this, we convert the packet to an $i$-packet with probability
$\nu_{i,j,k}/\nu$, setting it to the corresponding macro-atom state, and to 
a $k$-packet otherwise.

To speed up the calculations, we introduced an optional initial grey
approximation in the code. Controlled by two input parameters 
$\tau_\mathrm{Th,min}$ and $N_\mathrm{grey}$, the propagation of 
\textit{UVOIR} radiation in cells which have a Thomson optical depth 
  $\tau_\mathrm{Th}>\tau_\mathrm{Th,min}$
is performed including Thomson scattering as the only source of opacity.
If $\tau_\mathrm{Th}$ falls below $\tau_\mathrm{Th,min}$, and for all 
time steps later than $N_\mathrm{grey}$, we switch back to the non-grey 
treatment. This speeds up the initial phase of a calculation by a factor 
of $\sim10$, since much of the ejecta are still rather dense and photons 
cannot propagate. With a good selection of $\tau_\mathrm{Th,min}$ and 
$N_\mathrm{grey}$ the differences in spectra and light curves from a fully 
non-grey calculation are negligible.

\subsubsection{Extraction of spectra and light curves}
When the simulation has finished, we extract the spectral evolution by 
binning the escaping $r$-packets in frequency, time and direction of 
escape, accounting for light travel-time effects. Colour light curves 
are extracted from the spectral evolution by integrating the spectra 
over the appropriate filter functions \citep{Bessell1988,Bessell1990}. 
\textit{UVOIR} bolometric light curves are extracted by binning the 
escaping $r$-packets by time and angle. Similarly $\gamma$-ray spectra
and light curves can be obtained from the $\gamma$-packets.

Using the formal integral method developed by \citet{Lucy1999b} to extract 
spectra and light curves would substantially reduce the Monte Carlo noise.
However, the need to store line and continuum source functions in all grid 
cells for all lines and continua included in the simulation (typically 
$10^5$\dots$10^7$) causes the simulation memory requirements to be so high 
that this method becomes computationally prohibitive, even with current 
state-of-the-art super computers for 3D grids. Therefore all results 
presented here have been obtained using the direct binning approach.

\subsection{Plasma conditions}
To calculate opacities we need to know atomic level populations which in 
turn depend on the radiation field. In principle, these could be extracted 
exactly out of our simulation, but it is computationally restrictive to
record the complete set of level-by-level radiative rates for every grid
cell which would be required to do this. Therefore we 
use an approximate model for the radiation field and for excitation 
and ionisation conditions. Two different descriptions have been implemented 
in our code. The first parameterises the ionisation balance in a simple 
one-temperature description using the Saha formula. This is less demanding
of computational resources but also less accurate than our second which
treats ionisation in detail and simultaneously solves the thermal balance
equation.

\subsubsection{Radiation field models}
\label{sec:radiationfieldmodel}
Following \citet{Lucy2003}, the zeroth moment of the radiation field in 
a grid cell can be reconstructed from the Monte Carlo packets using the 
estimator 
  \begin{equation}
  \label{eq:Jnu_estimator}
  J_{\nu}\mathrm{d}\nu = 
    \frac{1}{4\pi\Delta t V} \sum_{\mathrm{d}\nu}\epsilon_{\nu}^{\mathrm{cmf}}\mathrm{d}s
  \end{equation}
where $\Delta t$ is the length of the time step, $V$ the grid cell volume, $\mathrm{d}s$ 
the trajectory length within the grid cell and $\epsilon_{\nu}^{\mathrm{cmf}}$ 
the packet energy in the cmf. The summation runs over all trajectory segments 
$\mathrm{d}s$ in $V$ for which the packet frequency $\nu\in\left[\nu,\nu+\mathrm{d}\nu\right]$.

As discussed above, we do not use this general description directly but 
parameterise the radiation field. For the simple one-temperature description 
of ionisation, the radiation field is parameterised as a black body,
  \begin{equation}
  J_{\nu}=B_{\nu}\left(T_{\mathrm{J}}\right)
  \label{eq:bbradfield}
  \end{equation}
at a temperature $T_\mathrm{J}$ corresponding to the local energy density 
of the radiation field ($B_\nu(T)$ is the Planck function). The parameter
$T_\mathrm{J}$ can be extracted by equating the frequency integrated 
estimator $\left\langle J\right\rangle \equiv\int_{0}^{\infty}J_{\nu}\,\mathrm{d}\nu$ 
(where $J_\nu$ is given by Equation \ref{eq:Jnu_estimator}) with the 
Stefan-Boltzmann law 
  \begin{equation}
  \label{eq:T_J}
  T_\mathrm{J}=\left( \pi\left\langle J\right\rangle /\sigma\right) ^{1/4}.
  \end{equation}
In our simple ionisation treatment, we also equate the kinetic temperature
$T_\mathrm{e}=T_\mathrm{J}$.

For the detailed ionisation treatment we parameterise the radiation field 
using a nebular approximation
  \begin{equation}
  \label{eq:nebularradfield}
  J_{\nu}=WB_{\nu}\left(T_{\mathrm{R}}\right),
  \end{equation}
with the radiation temperature $T_{\mathrm{R}}$ and the dilution factor $W$
as parameters. $T_{\mathrm{R}}$ is chosen such that the mean-intensity weighted
mean frequency of the radiation field 
  $
  \left\langle \nu \right\rangle \equiv 
    \int_{0}^{\infty}\nu J_{\nu}\,\mathrm{d}\nu/\int_{0}^{\infty} J_{\nu}\,\mathrm{d}\nu
  $
matches that of a blackbody at $T_{\mathrm{R}}$. Following 
\cite{Mazzali1993} $T_{\mathrm{R}}$ and $W$ can be extracted from 
$\left\langle J\right\rangle$ and $\left\langle \nu \right\rangle$ 
by 
  \begin{equation}
  W=\frac{\pi\left\langle J\right\rangle}{\sigma T_{\mathrm{R}}^{4}}
  \qquad\mbox{and}\qquad
  T_{\mathrm{R}}=\frac{h\left\langle \nu\right\rangle }{xk_{\mathrm{B}}}
  \label{eq:T_R}
  \end{equation}
with $x=360\cdot\zeta\left(5\right)/\pi^4$ and $\zeta\left(x\right)$ being Riemann's
$\zeta$-function. $T_\mathrm{J}$ remains defined as in the simple ionisation 
treatment (implying $T_{\mathrm{J}}=W^{1/4} T_{\mathrm{R}}$ from Equation 
\ref{eq:T_J}).

\subsubsection{Excitation and ionisation}
\label{sec:excandion}
Independent of the ionisation description used, we apply the Boltzmann 
formula evaluated at $T_\mathrm{J}$ 
  \begin{equation}
  \frac{n_{i,j,k}}{n_{0,j,k}}=\frac{g_{i,j,k}}{g_{0,j,k}}\, \exp\left( -\frac{\epsilon_{i,j,k}-\epsilon_{0,j,k}}{k_{\mathrm{B}}T_{\mathrm{J}}}\right) 
  \label{eq:boltzmann}
  \end{equation}
to calculate the population $n_{i,j,k}$ of a level $i$ of ion $j$ of 
element $k$ relative to the ion ground state population $n_{0,j,k}$. 
$g_{i,j,k}$ and $\epsilon_{i,j,k}$ are statistical weights and energies 
of the corresponding levels.

Assuming ionisation equilibrium in a radiation dominated 
environment\footnote{Currently we consider only photoionisation and 
radiative recombination. Temperatures in the recombination zones of 
iron group elements are sufficiently low that dielectronic recombination 
(calculated according to \citealt{Shull1982}) is small compared to 
radiative recombination. However we note that in high temperature 
environments it may become important, especially for intermediate mass 
elements. For more detailed studies our treatment could readily be extended 
to incorporate additional ionisation/recombination processes (including 
dielectronic recombination and non-thermal ionisation).}, we require
  \begin{equation}
  \sum_{i=0}^{\mathcal{N}_{j,k}}n_{i,j,k}\gamma_{i,j,k} =n_{e}N_{j+1,k}\sum_{i=0}^{\mathcal{N}_{j,k}}\left(\alpha_{i,j,k}^{\mathrm{sp}}+\alpha_{i,j,k}^{\mathrm{st}}\right)
  \label{eq:basicion}
  \end{equation}
where $\mathcal{N}_{j,k}$ is the number of levels associated with ion 
$j$ of element $k$, and $n_\mathrm{e}$ the density of free electrons. 
$\gamma_{i,j,k}$ denotes the photoionisation rate coefficient from a 
bound level $i,j,k$ and $\alpha^\mathrm{sp}_{i,j,k}$ is the rate 
coefficient for spontaneous recombination from the $(j+1)$-th ion to 
the bound state $i,j,k$ (see e.g. \citealt{Mihalas1978}). In our simple 
one-temperature ionisation treatment this reduces to the Saha equation
  \begin{equation}
  \label{eq:simpleion}
  \frac{N_{j,k}}{N_{j+1,k}n_{e}}=
\frac{U_{j,k}}{U_{j+1,k}}\frac{C}{T_{\mathrm{e}}^{3/2}} \,\exp\left(\frac{\epsilon_{0,j+1,k}-\epsilon_{0,j,k}}{k_{\mathrm{B}}T_{\mathrm{e}}}\right)
  \end{equation}
where 
  \begin{equation}
    \label{eq:sahaconst}
    C=\frac{1}{2}\left(\frac{h^2}{2\pi m_\mathrm{e}k_\mathrm{B}}\right)^{3/2}
  \end{equation}
and $U_{j,k}$ is the partition function of ion $j$ of element $k$.

For the detailed ionisation treatment we neglect stimulated recombination,
which is small compared to spontaneous recombination, and use the partition 
function together with
  \begin{equation}
  \label{eq:AlphaGamma}
  \Gamma_{j,k}\equiv\sum_{i=0}^{\mathcal{N}_{j,k}}\frac{n_{i,j,k}\gamma_{i,j,k}}{n_{0,j,k}}
  \qquad\mbox{and}\qquad
A_{j,k}^{\mathrm{sp}}\equiv\sum_{i=0}^{\mathcal{N}_{j,k}}\alpha_{i,j,k}^{\mathrm{sp}}
  \end{equation}
to obtain the following ionisation formula from Equation \ref{eq:basicion}
  \begin{equation}
  \label{eq:detailedion}
  \frac{N_{j,k}}{N_{j+1,k} n_{\mathrm{e}}}=\frac{A_{j,k}^{\mathrm{sp}}}{\Gamma_{j,k}}\cdot\frac{U_{j,k}}{g_{0,j,k}}
  \end{equation}
Since $\alpha_{i,j,k}^\mathrm{sp}$ does not depend directly on the radiation 
field, we calculate $A_{j,k}$ by summing all the $\alpha_{i,j,k}^\mathrm{sp}$
of the ion $j,k$ evaluated at the local kinetic temperature $T_\mathrm{e}$.

In contrast, the photoionisation rate coefficients $\gamma_{i,j,k}$ depend
directly on the radiation field and should be determined from volume-based
Monte Carlo estimators for most accurate results. However, we cannot afford 
to store $\gamma_{i,j,k}$ for each bound-free continuum in every grid cell.
Instead we derive $\gamma^*_{i,j,k}$ from an integration of the radiation 
field model (Equation \ref{eq:nebularradfield}) using the local 
values of $T_\mathrm{R}$ and $W$. Estimated in this way $\gamma^*_{i,j,k}$ 
is expected to be reliable for bound-free absorptions which lie around the 
peak of the radiation field. However, the radiation field model is inadequate 
at very blue frequencies where the spectrum is systematically affected by 
bound-free edges. Thus to track the effect of absorption continua on the 
radiation field, we record estimators $\gamma_{0,j,k}$ for the photoionisation 
rate coefficient of the ground level continua and use them to derive 
renormalisation coefficients $\zeta_{0,j,k}=\gamma_{0,j,k}/\gamma^*_{0,j,k}$. 
To obtain the $\gamma_{i,j,k}$ in $\Gamma_{j,k}$ of Equation \ref{eq:AlphaGamma} 
we renormalise all the integrated photoionisation rate coefficients via
  \begin{equation}
    \label{eq:gammarenorm}
    \gamma_{i,j,k} = \zeta_{0,j',k'}\gamma^*_{i,j,k},
  \end{equation}
where we take the $\zeta_{0,j',k'}$-value from the ground level continuum
which lies closest in edge frequency to the considered continuum (i.e., the 
one for level $i,j,k$). For continua redder than the reddest ground level 
continuum, $\zeta$ is set to 1.

For the zeroth time step, and also for cells treated in our initial grey
approximation (see Section \ref{sec:rpackets}), we use the simple ionisation 
treatment. We also assume that at all times prior to our zeroth time step 
the ejecta are sufficiently optically thick, that all the radiation is trapped. 
Thus we calculate initial temperatures $T_{\mathrm{e},0}=T_{\mathrm{R},0}=T_{\mathrm{J},0}$ 
for each grid cell from the energy released by the decay of $^{56}$Ni 
in the cell prior to the midpoint of the zeroth time step. Since the 
trapping is not really perfect, it takes a few (typically $\sim5$) time 
steps until the propagation of the energy packets washes out the effect 
of this boundary condition on temperatures and level/ion populations. 

In practise this boundary condition primarily effects calculations using the 
detailed ionisation description. In this case, it is more important to have 
accurate level populations as errors can feed back to the thermal balance
calculation. Thus we typically do the first $\sim 10$ time steps of a 
calculation in the simple ionisation treatment to obtain reliable initial 
conditions for the detailed ionisation treatment at all later times. Since
our simple ionisation treatment reproduces LTE conditions we expect it to be
a good approximation at early times when optical depths are high.

\subsubsection{Thermal balance}
In the detailed ionisation description, we also solve for the local kinetic 
temperature $T_\mathrm{e}$ in a grid cell by balancing the local heating 
$\mathcal{H}$ and cooling $\mathcal{C}$ rates. The heating rates are 
calculated from the previous time step using volume-based Monte Carlo 
estimators for the heating by $\gamma$-rays \citep[see][]{Lucy2005} 
and free-free absorptions \citep[see][]{Lucy2003} and event-based Monte 
Carlo estimators for the heating by collisional deexcitation/recombination. 

We cannot afford to store a bound-free heating estimator for each continuum
in every grid cell. Thus bound-free heating rates 
  \begin{equation}
  \label{eq:bfheating}
  \mathcal{H}^\mathrm{bf}_{i,j,k}=
    n_{i,j,k}h^\mathrm{bf}_{i,j,k}=
    n_{i,j,k}\left(\gamma^\mathrm{E}_{i,j,k}-\gamma_{i,j,k}\right)h\nu_{i,j,k},
  \end{equation}
\citep[with $\gamma^\mathrm{E}_{i,j,k}$ the modified rate coefficient for 
photoionisation as defined by][]{Lucy2003} are not obtained directly from 
Monte Carlo estimators but from an integration of the radiation field model 
using the local values of $T_\mathrm{R}$ and $W$ yielding $h^\mathrm{bf,*}_{i,j,k}$. 
As for the photoionisation rate coefficients (Section \ref{sec:excandion}), 
we then use estimators $h^\mathrm{bf}_{0,j,k}$ for the bound-free heating 
coefficient of the ground level continua to derive renormalisation coefficients
$\xi_{0,j,k}=h^\mathrm{bf}_{0,j,k}/h^\mathrm{bf,*}_{0,j,k}$. These are used 
to calculate the bound-free heating coefficients
  \begin{equation}
  h^\mathrm{bf}_{i,j,k}=\xi_{0,j',k'}h^\mathrm{bf,*}_{i,j,k}
  \end{equation}
where we again take the $\xi_{0,j',k'}$ which lies closest in frequency to 
the considered continuum (i.e., the one for level $i,j,k$) to track the 
effect of absorption continua on the radiation field. Continua which are 
redder than the reddest ground-level continuum are not renormalised.

We include cooling by free-free and free-bound emission, collisional 
excitation/ionisation and adiabatic expansion. Since these cooling rates 
do not depend on the radiation field directly, they are calculated from the 
current set of level populations and $T_\mathrm{e}$ as described in 
\citet{Lucy2003}. The adiabatic cooling rate is given by
  \begin{equation}
  \mathcal{C}^\mathrm{ad}=p\frac{\mathrm{d}V}{V}
  \label{eq:adcool}
  \end{equation}
with $p$ the gas pressure and $V$ the volume of the grid cell.

Balancing the heating and cooling rates on a temperature interval 
$\left[T_\mathrm{min},T_\mathrm{max}\right]$ with $T_\mathrm{min}=3500\,\mathrm{K}$
and $T_\mathrm{max}=1.4\cdot10^5\,\mathrm{K}$ in our simulations, leads to a new 
value of $T_\mathrm{e}$. For cells in which the heating and cooling rates 
cannot be balanced on this interval, $T_\mathrm{e}$ is set to the upper/lower
boundary value if the heating/cooling rates dominate. The same temperature limits
apply to $T_\mathrm{J}$ and $T_\mathrm{R}$. $T_\mathrm{max}$ is chosen such that 
it will be only reached at the very earliest times in the inner iron-rich core
where the opacity is so high that the radiation is trapped (actually for the 
simulations described in this paper $T_\mathrm{max}$ was never reached). 
$T_\mathrm{min}$ will be reached at late times but only in layers which are
sufficiently optically thin that they have no strong influence on the spectrum
formation.

Assuming that time steps are short enough and the radiation field does not
depend strongly on the evolution of $T_\mathrm{e}$ we use the obtained 
value for $T_\mathrm{e}$ and the same Monte Carlo estimators as before to 
re-solve for the populations and then for thermal balance again. Repeating 
this iteratively we reach a converged solution for $T_\mathrm{e}$ and the 
ionisation state. This iteration can be extended to an outer loop in which 
the Monte Carlo simulation is also repeated. \citet[][Section 3.4]{Kasen2006} 
showed that, even from crude initial conditions, such an iteration converges 
very rapidly. By using short time steps and initial thermodynamic quantities 
from the previous time step, we found that such an outer iteration of the 
Monte Carlo experiment does not affect the results strongly. Thus we do not 
iterate on the Monte Carlo simulation for each time step in the simulations 
described below (however the option to do so remains in the code).

\subsection{Model atoms}
\label{sec:modelatoms}
The most important opacity source in supernova envelopes is the wealth of
bound-bound line transitions associated with the iron group elements 
synthesised by the thermonuclear explosion. Thus an extensive line list 
which covers all the important species is needed. We restrict our atomic
data to the lowest five ionisation stages of elements up to Zn, neglecting
H, Li, Be and B. To study the sensitivity to the choice of atomic data, 
we use two datasets which are based on the $\sim 5\cdot10^5$ lines of 
\citet{Kurucz1995} (CD23) and the more modern theoretical atomic data 
computed by \citet{Kurucz2006}\footnote{Available at 
\tt{http://kurucz.harvard.edu/atoms.html}.} which are far more 
comprehensive but still not complete. For example this dataset contains 
$\sim 3.5\cdot10^7$ lines associated with the important second and third 
ions of Fe, Co and Ni compared to only $\sim1.2\cdot10^5$ in CD23.

Since it is computationally too demanding to use all of the lines in 
a dataset, when constructing model atoms we apply cuts in 
$\log\left(gf\right)$ to reduce the number of lines. From the comprehensive
dataset we take only the lines of Fe\,{\sc ii}, Fe\,{\sc iii}, Co\,{\sc ii},
Co\,{\sc iii}, Ni\,{\sc ii} and Ni\,{\sc iii}, the most important species in 
the spectrum forming region, and continue to use the data of CD23 for all 
the other species and remaining ions of Fe, Co and Ni. In this way, we 
created a set of model atoms (summarised in Table \ref{tab:modelatoms}) 
which are used in the following calculations.

Bound-free cross sections are obtained from the fits by \citet{Verner1995}
or \citet{Verner1996} where available. A hydrogenic approximation is 
adopted for excited configurations.

\begin{table}
  \centering
  \caption{Model atoms used in the calculations. Source A is \citet{Kurucz1995}, 
  Source B \citet{Kurucz2006}.}
  \label{tab:modelatoms}
  \begin{tabular}{@{}llccc@{}}
  \hline
  Model name & Source & $\log(gf)$ & Levels & Lines (total)\\
  \hline
  cd23\_gf-2 & A & -2   & 18237 & $1.4\cdot10^5$\\
  cd23\_gf-3 & A & -3   & 18815 & $2.5\cdot10^5$\\
  cd23\_gf-5 & A & -5   & 19472 & $4.1\cdot10^5$\\
  cd23\_gf-20 & A & -20 & 21100 & $4.7\cdot10^5$\\
  big\_gf-3 & B (Fe\,{\sc ii--iii}) & -3 & 41829 & $3.5\cdot10^6$\\
            & B (Co\,{\sc ii--iii}) & -3 & 31816 &\\
            & B (Ni\,{\sc ii--iii}) & -3 & 24853 &\\
            & A (other) & -3        & 16616 &\\
  big\_gf-4 & B (Fe\,{\sc ii--iii}) & -4 & 47141 & $8.2\cdot10^6$\\
            & B (Co\,{\sc ii--iii}) & -4 & 36442 &\\
            & B (Ni\,{\sc ii--iii}) & -4 & 28022 &\\
            & A (other) & -4        & 17523&\\
  \hline
  \end{tabular}

\end{table}

\section{Application to the W7 model}
\label{sec:w7}
As a first test we calculate the spectral evolution for the one-dimensional 
deflagration model W7 \citep{Nomoto1984,Thielemann1986} which has already been investigated
in several other radiative transfer studies 
\citep[e.g.][]{Jeffery1992,Hoeflich1995,Nugent1997,Lentz2001,Salvo2001,Baron2006,Kasen2006}
and is found to be in good overall agreement with observations. 

\begin{figure}
  \centering
  \includegraphics[angle=90]{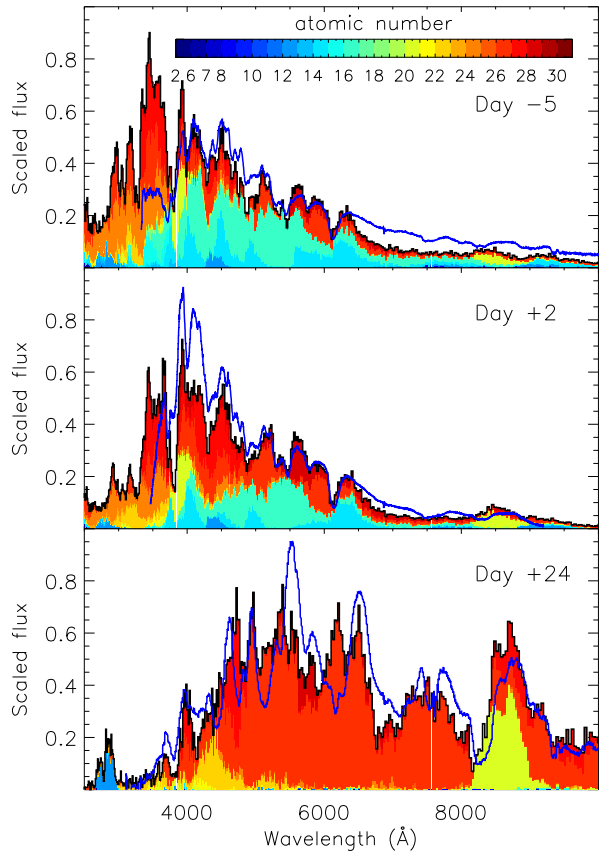}
  \caption{Spectra for the W7 explosion model at 5 days before and 2 and 24 
    days after maximum light in $B$ band (from top to bottom) calculated using 
    the detailed ionisation description and atomic dataset cd23\_gf-5. Overplotted 
    in blue are the observed spectra of SN 1994D for comparison (no redshift
    and extinction corrections have been applied). The colour coding shows the 
    element associated with the last line emission of escaping photons. Atomic
    numbers in the colour legend are associated with the colour right to the
    number centre.}
  \label{fig:w7spectra}
\end{figure}

Our calculations use $5\cdot 10^6$ energy packets to follow the 
spectral evolution over 100 time steps from 2 to 80 days after explosion ($\Delta\log(t)\sim0.037$) and 
were performed on a $50^3$ grid using the cd23\_gf-5 model atom (see Table 
\ref{tab:modelatoms} for details) with our detailed ionisation treatment.
The first 10 time steps have been calculated in the simple ionisation treatment
to obtain reliable initial values (see the discussion in Section \ref{sec:excandion}).
To save computational time we make use of our initial grey approximation setting $\tau_\mathrm{Th,min}=15 $ and $N_\mathrm{grey}=40$. 

Figure \ref{fig:w7spectra} shows the 2500 to 10\,000 $\mathrm{\AA}$ spectra 
for 15, 22 and 44 days after explosion, corresponding to -5, +2 and +24 days
relative to $B$ band maximum. The colour coding shows the element associated
with the last line emission of escaping photons, giving an indication of the 
elements responsible for spectral features (absorption features of individually 
strong lines can be identified in the plot from their associated P Cyg emission). 
The contribution of photons which escaped after a bound-free or free-free emission 
to the total flux is negligible ($<10^{-3}$) and not shown. The most striking 
individual features are the Ca\,{\sc ii} near-infrared (NIR) triplet at 
$\sim 8600\,\mathrm{\AA}$ and the characteristic Si\,{\sc ii} line at 
$6355\,\mathrm{\AA}$ which is clearly visible around maximum light. 
While the outer layers, which are dominated by intermediate mass elements, 
are optically thick, the spectra are dominated by these elements. 
Around maximum light the outer layers become optically thin and the spectra 
start to be dominated by Fe group elements which completely take over at 
later times. Only individual strong lines of the lower mass elements (e.g. 
the Ca\,{\sc ii} NIR triplet) persist.

Overplotted with our synthetic spectra are observations of SN 1994D \citep{Patat1996}
for the corresponding epochs. Given that the W7 model has not been tuned to
any particular supernova, the agreement of our model spectra with the observed 
ones is good. We reproduce the main spectral features, e.g. the Si\,{\sc ii} 
line and the Ca\,{\sc ii} NIR triplet, as well as the overall flux
distribution. However there are obvious differences: at early times our model
spectra have a strong excess in the UV flux below 4000\,$\mathrm{\AA}$
where Fe\,{\sc ii}, Fe\,{\sc iii}, Co\,{\sc iii} and Co\,{\sc iv} dominate
the spectrum formation. At late times we obtain emission in Fe\,{\sc ii} at 
$\sim6050\,\mathrm{\AA}$ where the data shows an absorption feature. Model 
spectra calculated for W7 by \citet{Kasen2006} show the same discrepancies 
compared to SN 1994D. This could be an indication that some details of the 
explosion model would need to be changed to obtain better agreement with 
SN 1994D.

\begin{figure*}
  \centering
  \includegraphics[angle=90]{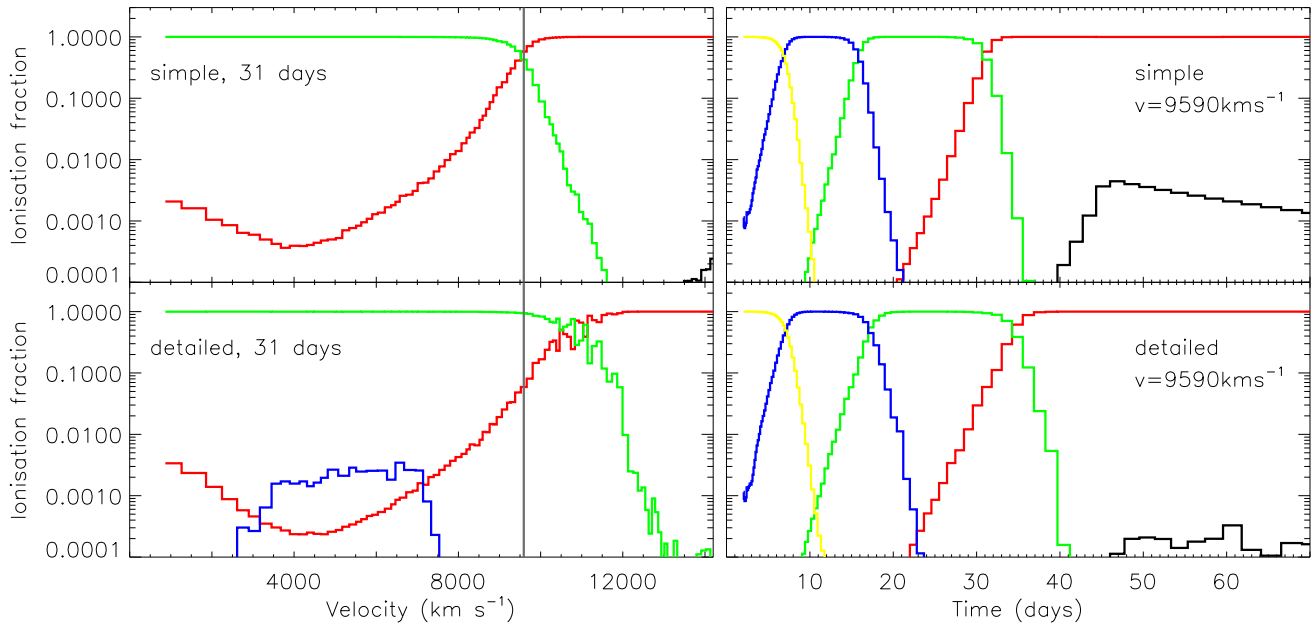}
  \caption{The left panels show the radial ionisation structure of Fe in
    the W7 explosion model at 31 days after the explosion obtained with
    the simple (top) and detailed ionisation treatment (bottom) and model
    atom cd23\_gf-5. Black/red/green/blue/yellow lines represent 
    Fe\,{\sc i}/{\sc ii}/{\sc iii}/{\sc iv}/{\sc v}. The outer layers of the
    model, which consist only of unburned material and extend up to 
    $22\,800\,\mathrm{km\,s}^{-1}$, are not shown. 
    The right panels show the time evolution of the Fe ionisation structure
    at a velocity of $9590\,\mathrm{km\,s}^{-1}$ (indicated by the grey vertical 
    line in the left panels) from 2 to 70 days for the two different ionisation
    treatments.}
  \label{fig:w7ionbalance}
\end{figure*}

\subsection{Simple versus detailed ionisation treatment}
\label{sec:w7_simplevsdetailed_ion}

In the following we investigate how the two different ionisation treatments
(see Section \ref{sec:excandion}) which are implemented in our code affect 
the spectral evolution of a given explosion model. Therefore we also followed 
the spectral evolution of the W7 model using the cd23\_gf-5 atomic data from 
2 to 80 days after the explosion using the simple ionisation treatment.

Since differences in the spectral evolution must originate from the 
different ionisation treatment, we compare ionisation fractions from 
our different simulations in Figure \ref{fig:w7ionbalance}. We focus 
on Fe since the iron group elements, which have similar ionisation 
structure to each other, dominate the spectra for the relevant epochs. 
The left panels of Figure \ref{fig:w7ionbalance} show the ionisation fractions
as a function of radial velocity at 31 days after the explosion for the simple 
and detailed ionisation treatment. The right panels show the same ionisation 
fractions as a function of time at a radial velocity of $9590\,\mathrm{km\,s}^{-1}$,
which is marked by the grey vertical line in the left panels. This is the velocity
at which the iron group mass fraction of the ejecta drops to 0.5 and thus is
around the outer velocity of the iron-rich inner core.

\begin{figure}
  \centering
  \includegraphics[angle=90]{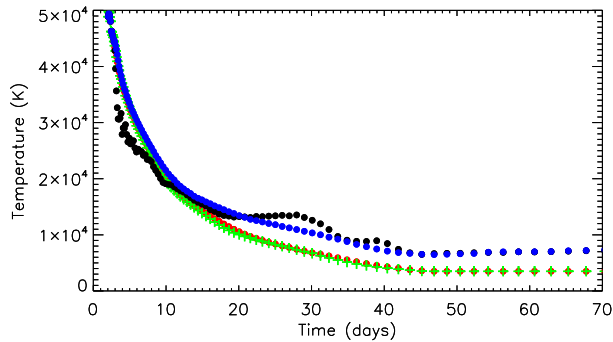}
  \caption{Temperature evolution at a radial velocity of $9590\,\mathrm{km\,s}^{-1}$
    i.e. at the outer edge of the iron-rich core. The green crosses show 
    $T_\mathrm{J}$ in the simple ionisation treatment. The circles are the 
    three temperatures used in the detailed ionisation treatment and show
    the kinetic temperature $T_\mathrm{e}$ (black), the radiation
    temperature $T_\mathrm{R}$ (blue) and $T_\mathrm{J}$ (red). At 
    around 45 days, $T_\mathrm{J}$ hits the lower temperature boundary
    of $3500\,\mathrm{K}$ and is not allowed to drop further.}
  \label{fig:Ttime}
\end{figure}

From the left panels we see that, using the detailed ionisation treatment, the 
ejecta are more highly ionised at higher radial velocities. The right panels 
show that, with the detailed ionisation treatment, the ejecta also
stay more highly ionised for a longer time at a given radius and never
recombine as fully as with the simple ionisation treatment at late times.
In the simple ionisation treatment, the ionisation is determined only by 
the energy density of the radiation field calculated from the $r$-packets.
In the detailed ionisation treatment, the frequency distribution of
the calculated radiation field has an important effect. 
$T_\mathrm{R} > T_\mathrm{J}$ and $W < 1$ (see Equations 6 -- 8) with 
$T_\mathrm{R}$ and $W$ departing more strongly from $T_\mathrm{J}$ and 1, 
respectively, with increasing time and decreasing ejecta density and 
opacity. Thus, there tend to be more high-energy ionising photons than 
suggested by the radiation field model of the simple ionisation treatment. 
This gives rise to higher ionisation states, particularly for later times. 
This is illustrated in Figure~\ref{fig:Ttime} which shows the time evolution 
of the radiation temperature $T_\mathrm{R}$ and the temperature $T_\mathrm{J}$ 
corresponding to the energy density of the radiation field at a radial velocity 
of $9590\,\mathrm{km\,s}^{-1}$. While $T_\mathrm{J}$ drops similarly for the 
simple and detailed ionisation treatment, $T_\mathrm{R}$ for the detailed 
ionisation treatment stays significantly hotter from 20 days on. 

$T_\mathrm{e}$ shows a similar decline and stays somewhat below $T_\mathrm{R}$ 
up to about maximum light (note that at very early times, prior to $\sim 10$ days,
this region is dominated by Fe\,{\sc v} for which line transitions were not
included -- this likely explains the less complete coupling of $T_\mathrm{e}$
to the radiation field at these early epochs).
Afterwards $\gamma$-ray heating dominates the total heating rate and controls 
the kinetic temperature. Although the $\gamma$-ray heating decreases smoothly 
with time, $T_\mathrm{e}$ shows distinct changes. These arise when recombination 
from one dominant ion to the next occurs and the contributing cooling processes 
change abruptly. In order to retain the balance between the heating and cooling 
rates then $T_\mathrm{e}$ must change rapidly.

\begin{figure}
  \centering
  \includegraphics[angle=90]{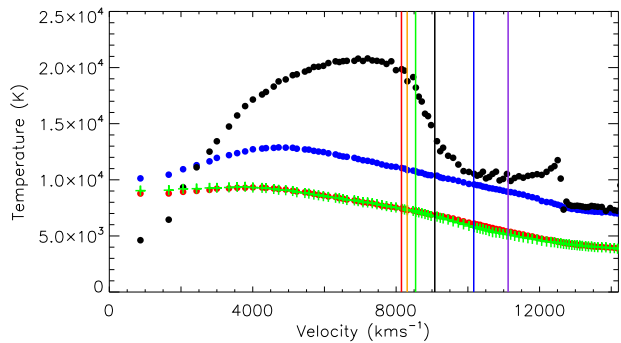}
  \caption{Radial temperature distribution at 31 days after the explosion.
    The outer layers of the model, which consist only of unburned material 
    and extend up to $22\,800\,\mathrm{km\,s}^{-1}$, are not shown. The 
    green crosses show $T_\mathrm{J}$ in the simple ionisation treatment.
    The circles are the three temperatures used in the detailed ionisation 
    treatment and show the kinetic temperature $T_\mathrm{e}$ in black, 
    the radiation temperature $T_\mathrm{R}$ in blue and $T_\mathrm{J}$ 
    in red. The black/violet/blue/green/red/orange vertical lines show 
    the mean radii of last scattering in bolometric/$U$/$B$/$V$/$R$/$I$ light.
    The sharp drop in $T_\mathrm{e}$ around $12\,500\,\mathrm{km\,s}^{-1}$
    is associated with a strong change in composition in the W7 model.
    The iron group abundance abruptly drops at this point and 
    intermediate-mass elements dominate.}
  \label{fig:T_radial}
\end{figure}

Figure~\ref{fig:T_radial} illustrates the radial temperature distribution at 
31 days after the explosion. $T_\mathrm{J}$ for the two different ionisation
treatments shows no strong difference. However $T_\mathrm{R}$ in the detailed 
ionisation treatment differs strongly from $T_\mathrm{J}$ over the entire 
ejecta, indicating that the radiation field is bluer and more dilute that
would be implied by our radiation field model in the simple ionisation 
treatment. The kinetic temperature is mainly controlled by $\gamma$-ray 
heating which dominates over the other heating rates outside $\sim2500\,
\mathrm{km\,s}^{-1}$ and varies only gradually with velocity, declining 
outwards. This leads to $T_\mathrm{e}>T_\mathrm{R}$. Around recombination
fronts the contributing cooling processes change abruptly. This means 
that $T_\mathrm{e}$ has to change rapidly in such regions in order that 
the heating rate remains balanced by the total cooling rate and leads to 
the rapid decline in $T_\mathrm{e}$ between 8000 and 10\,000 km/s where
the recombination from Fe\,{\sc iii} to Fe\,{\sc ii} happens (compare Figure~\ref{fig:w7ionbalance}).
The vertical lines show the mean radii of last scattering for packets which had 
their last scattering at $\sim31$ days. Because of line blocking the optical depths
are biggest in the UV and blue and the different bands probe deeper
into the ejecta in a sequence from blue to red. An exception is the mean radius
of last scattering in the $I$ band which lies at $\sim8500\,\mathrm{km\,s}^{-1}$,
between the $R$ and $V$ bands. This is because the $I$ band has significant 
contribution from the Ca\,{\sc ii} NIR triplet which forms outside the iron-rich
core.

\begin{figure*}
  \centering
  \includegraphics[angle=90]{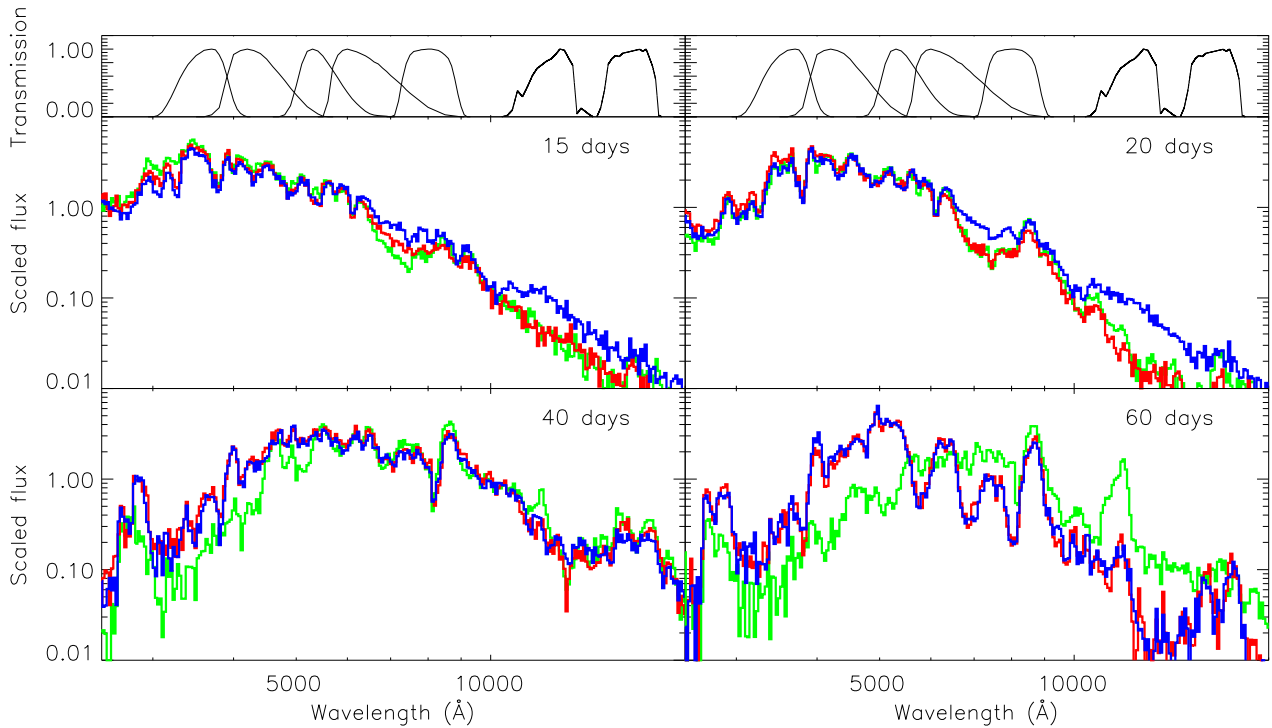}
  \caption{Spectra for the W7 explosion model at 15, 20, 40 and 60 days 
    after the explosion, respectively. The green and red lines were 
    calculated with the cd23\_gf-5 atomic dataset using the simple and 
    detailed ionisation treatment, respectively. The blue line was 
    calculated with the big\_gf-4 atomic dataset and the detailed ionisation 
    treatment. The top panels show (from left to right) the normalised 
    $U$,$B$,$V$,$R$,$I$,$J$ and $H$ passbands of \citet{Bessell1990} 
    and \citet{Bessell1988}.}
  \label{fig:comparew7spectra}
\end{figure*}

That the ejecta stay more ionised for longer in the detailed ionisation 
description has a direct influence on the spectra at late times. This is 
illustrated in Figure \ref{fig:comparew7spectra} which shows spectra
for 15, 20, 40 and 60 days after the explosion for the simple and detailed 
ionisation treatments. While there are no differences up to about maximum 
light in $B$ band at 20 days, the spectra calculated using the detailed 
ionisation treatment are significantly bluer at late times. This is because 
they still have more Fe\,{\sc ii} contribution relative to Fe\,{\sc i} 
than the spectra using the simple ionisation treatment (Fe\,{\sc ii} 
has bluer lines). The simple ionisation treatment shifts more flux into 
the red and NIR leading to a strong Fe\,{\sc i} emission feature at 60 
days which is not seen in observations.

\begin{figure*}
  \centering
  \includegraphics[angle=90]{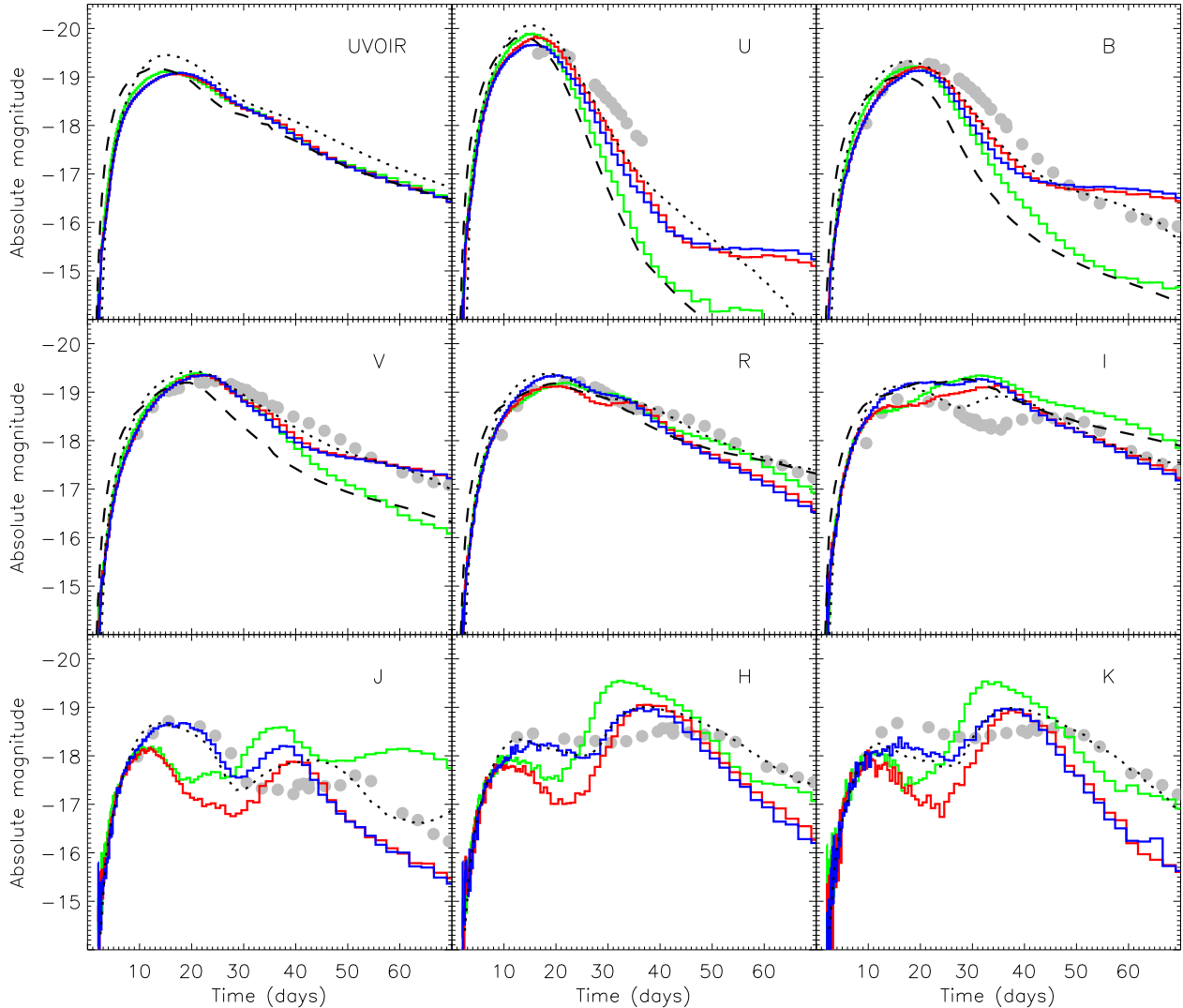}
  \caption{\textit{UVOIR} bolometric and \textit{U,B,V,R,I,J,H,K} light curves
    for the W7 explosion model. The green/red lines are the light curves 
    obtained with the simple/detailed ionisation treatment and atomic dataset 
    cd23\_gf-5. The blue lines are calculated with the detailed ionisation 
    treatment and atomic dataset big\_gf-4. The small fluctuations e.g. in the $U$ 
    light curve of the simple ionisation treatment and the NIR light curves 
    are due to Monte Carlo noise.
    For comparison observations of SN 2001el \citep{Krisciunas2003} 
    are overplotted as grey circles assuming a distance modulus of 31.54 
    and the reddening law of \citet{Cardelli1989} with $A_V=0.5$ and $R_V=3.1$.
    Furthermore W7 light curves obtained with the Monte Carlo radiative transfer 
    code \textsc{sedona} (Kasen priv. comm.; \textit{UVOIR},\textit{U,B,V,R,I,J,H,K} dotted 
    lines) and the radiation hydrodynamics code \textsc{stella} (Sorokina priv. comm.; 
    \textit{UVOIR},\textit{U,B,V,R,I} dashed lines) are shown.}
  \label{fig:w7lightcurves}
\end{figure*}

The same effect influences the broad band light curves calculated using 
the simple and detailed ionisation treatment which are shown in Figure 
\ref{fig:w7lightcurves}. The band passes of \citet{Bessell1990} and 
\citet{Bessell1988} are shown in the top panel of Figure \ref{fig:comparew7spectra} 
to clarify the following discussion. Like the spectra, the light curves
are rather similar before maximum light in $B$ band at $\sim 20$ days, 
with slight differences in the NIR bands. After maximum light there are 
strong differences. The $U$, $B$ and $V$ light curves calculated using 
the simple ionisation treatment fade much quicker than those using 
the detailed ionisation treatment. In contrast, $R$, $I$, $J$, $H$ and 
$K$ stay brighter with $J$ showing a third maximum which is associated 
with the appearance of the strong Fe\,{\sc i} emission feature in the 
bottom right panel of Figure \ref{fig:comparew7spectra} at about 60 days.

Of particular interest is the $R$ band which shows a clear secondary maximum
using the detailed ionisation treatment but only a slight plateau using the
simple ionisation treatment. As pointed out by \citet{Kasen2006a}, the
secondary maximum forms when the zone in which doubly ionised iron group 
elements recombine hits the inner iron-rich core. It is then that the 
redistribution of flux from the UV and blue part of the spectrum into 
the red and NIR by fluorescence is most effective. 
This argumentation is confirmed and explains the differences in the $R$ 
band light curve between the simple and the detailed ionisation treatment. 
In the right panels of Figure \ref{fig:w7ionbalance}, which shows the 
ionisation fractions of Fe as a function of time at the outer edge of the
iron-rich inner core, we see that for the simple ionisation treatment (top
panel) the transition from Fe\,{\sc iii} to Fe\,{\sc ii} happens at $\sim 33$
days while it occurs at $\sim 36$ days in the detailed ionisation treatment. 
These times correspond to the times of the secondary maximum of the $R$ 
band light curve in Figure \ref{fig:w7lightcurves}. Thus, with the simple 
ionisation treatment, the secondary maximum is blended in the first peak
such that it is not clearly evident in the light curves.

Taking into account all the effects which have been discussed in this section
we conclude that a detailed treatment of ionisation, consistent with the 
properties of the radiation field, has an important influence and is 
needed to obtain reliable broad-band light curves and spectra after 
maximum light. In particular, we note that the colour evolution obtained 
from calculations using a simple LTE description of ionisation is subject 
to strong uncertainties.

\subsection{Influence of atomic data}
\begin{figure}
  \centering
  \includegraphics[angle=90]{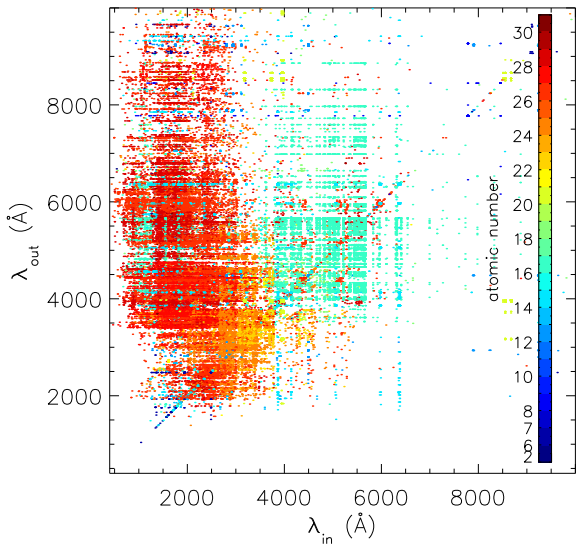}
  \caption{Wavelength redistribution in the last line transition of photon 
    packets which escaped at 20 days after the explosion. Each point displays
    the wavelength of an escaping photon packet versus the wavelength at which 
    the same packet had been absorbed immediately before emission. Thereby a
    single point can represent multiple photon packets which have undergone 
    the same absorption and emission processes before they escaped. Points
    on the diagonal line are photon packets which last undergone resonance
    scattering. Points above (below) that line have been redistributed into 
    the red (blue) by (reverse) fluorescence. The colour coding shows by which 
    element an escaping photon was last emitted (atomic numbers refer to the
    colour on top of the number centre) thus indicating that the iron
    group elements are most efficient in redistributing flux into the red
    by fluorescence. The data were extracted from our W7 calculation using the
    detailed ionisation treatment and the cd23\_gf-5 model atom.}
  \label{fig:redmat}
\end{figure}

As already mentioned, the vast number of line transitions associated with 
the iron group elements are the main contributors to the opacity. 
Furthermore the lines provide an efficient way to redistribute radiation 
from the ultraviolet and blue part of the spectrum into the red and NIR 
by fluorescence \citep{Pinto2000b}. This is illustrated in Figure 
\ref{fig:redmat} which shows, for 20 days after explosion, the wavelength 
at which escaping photon packets were emitted versus the wavelength at which 
the same packets had been last absorbed prior to escape (using atomic dataset 
cd23\_gf-5 and the detailed ionisation treatment). It shows that many 
absorptions in the UV are followed by emission in the red and NIR, but 
also that the reverse process happens too (see discussion by 
\citealt{Mazzali2000a,Sauer2008a}). From the colour coding we see that the 
iron group elements are most effective in this redistribution.

\begin{figure}
  \centering
  \includegraphics[angle=90]{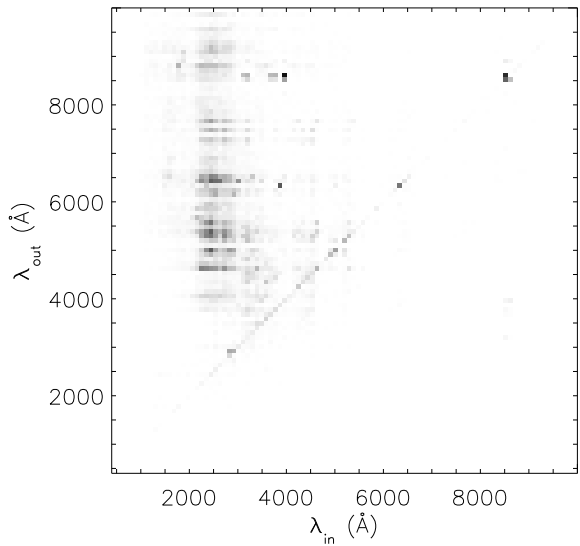}
  \caption{Wavelength redistribution in the last line transition of photon 
    packets which escaped at 35 days after the explosion. The escaping photon
    packets have been binned into a wavelength grid. The shade of a grid point
    indicates how many photon packets escaped in this bin (the darker the more
    packets escaped). Points on the diagonal line are photon packets which last 
    underwent resonance scattering. Points above (below) that line have been 
    redistributed into the red (blue) by (reverse) fluorescence. The dark dots
    at  $\lambda_\mathrm{out}\sim 8600\,\mathrm{\AA}$ and  $\sim 6300\,\mathrm{\AA}$ 
    show the flux redistribution from the Ca\,{\sc ii} H and K lines into the 
    Ca\,{\sc ii} NIR triplet and in Si\,{\sc ii} respectively. 
    The data were extracted from our W7 calculation using the detailed 
    ionisation treatment and the cd23\_gf-5 model atom.}
  \label{fig:redmat_lines}
\end{figure}

Figure \ref{fig:redmat_lines} shows the redistribution at 35 days after 
the explosion with escaping photon packets binned into a wavelength grid, 
and indicating how many packets escaped in a bin by its grey shade. Thus
individual strong lines become visible. The dark dots at $\lambda_\mathrm{out}\sim 8600\,\mathrm{\AA}$ are associated with emission in the Ca\,{\sc ii} NIR 
triplet which results from absorption in the Ca\,{\sc ii} H and K lines 
at $\lambda_\mathrm{in}\sim 4000\,\mathrm{\AA}$ as well as from resonance 
scattering in the NIR triplet itself. In a similar way the Si\,{\sc ii} 
line at $\sim 6355\,\mathrm{\AA}$ shows up as a resonance line and in 
fluorescence with photons absorbed at $\sim 4000\,\mathrm{\AA}$. This 
underlines that a fully detailed treatment of line formation is needed
to make reliable predictions of individual spectral features and to a lesser
extent also for the broad-band light curves. Especially the $I$ band light 
curve will be affected since it has a strong contribution from the Ca\,{\sc ii} 
NIR triplet (see top panels of Figure \ref{fig:comparew7spectra}). 

To simulate this redistribution properly an atomic dataset as complete as 
possible is desirable. Unfortunately the simulations become more and more 
expensive for bigger atomic datasets and the atomic data is not completely 
known. Thus in the following we study the effect of incomplete atomic 
data by using datasets of different completeness to calculate the 
spectral evolution of W7.

We started with the line list of \citet{Kurucz1995} and tried to reproduce
the results of the previous section using atomic datasets with cuts of -2 
and -3 in  $\log\left(gf\right)$ compared to -5 which was used so far (see 
Table \ref{tab:modelatoms} for details on the atomic data). The light curves
obtained with the -2 cut differ strongly from the previous ones, 
indicating that lines weaker than -2 in $\log\left(gf\right)$ play an 
important role. With the -3 cut, the light curves in the optical bands 
are almost identical but there are still some minor differences in the NIR 
bands. Going further to a -20 cut does not change the results compared to 
the -5 cut. So the cd23\_gf-5 atomic dataset, which was used above, is sufficient 
to simulate the redistribution which is possible within the line list of 
\citet{Kurucz1995}.

However, this line list with its meagre total of $\sim 5\cdot10^5$ lines,
contains only a tiny fraction of the millions of lines expected of the 
iron group elements. As many of these lines are associated with transitions
amongst highly excited levels, this line list lacks especially lines in 
the red and NIR parts of the spectrum. It was pointed out by \citet{Kasen2006} 
and \citet{Kasen2006a} that these missing lines are crucial to obtain light 
curves in good agreement with observations, especially in the NIR. To 
study this, we use atomic dataset big\_gf-4 (for details see Table \ref{tab:modelatoms} 
and Section \ref{sec:modelatoms}) which includes atomic data for 
Fe\,{\sc ii}, Fe\,{\sc iii}, Co\,{\sc ii}, Co\,{\sc iii}, Ni\,{\sc ii} 
and Ni\,{\sc iii} from more modern theoretical computations, increasing 
the number of lines by a factor of $\sim 20$ compared to the cd23\_gf-5 
atomic dataset.

The bolometric and broad-band light curves obtained for the W7 model using
the detailed ionisation treatment and atomic dataset big\_gf-4 are shown 
by the blue line in Figure \ref{fig:w7lightcurves}. $UVOIR$ bolometric and 
the $U$, $B$ and $V$  light curves are not strongly affected. In contrast, 
the enhanced redistribution of flux from the UV and blue to the red and NIR 
at early times, which can be seen comparing the red and blue curves in the top 
panels of Figure \ref{fig:comparew7spectra}, increases the first peak in the 
$R$, $I$, $J$, $H$ and $K$ bands and causes strong differences for
$t \lesssim 40$ days. As a consequence of this redistribution, the $U$ and
$B$ band light curves using the big\_gf-4 atomic dataset are slightly dimmer
between $\sim 10$ and $\sim 40$ days than those using the cd23\_gf-5 atomic 
dataset. At later times, the light curves are very similar.

The increase in the magnitude of the primary peak in the NIR brings the model 
light curves to better agreement with observations. The distinct secondary 
maximum which is present in the $R$ band light curve using the cd23\_gf-5 
atomic dataset changes -- as it is observed -- to a shoulder using atomic 
dataset big\_gf-4. However it occurs too early compared to the observations
and especially the $H$ 
and $K$ bands show systematic differences from the observations having
secondary peaks brighter than their primary peaks. This could be due to
a further lack of atomic data, which might increase the redistribution into 
the $H$ and $K$ bands. However, calculations with the big\_gf-3 atomic dataset
(see Table \ref{tab:modelatoms}), which has fewer lines by a factor of 
$\sim 2$, do not show significant differences around the first peaks, 
suggesting that very weak lines are not critical in this.
It is possible, however, that the atomic data set in use is still not
complete in terms of lines stronger than a -4 cut in $\log(gf)$ -- which
we cannot exclude -- or that more iron group elements or ions should be 
included from the comprehensive atomic data set \citep{Kurucz2006}. However 
memory limitations and computational costs become a severe issue if the 
atomic data are to be significantly expanded and we note that the small 
abundances of the other iron group elements suggest that they are less 
likely to have a major effect. Alternatively, the explosion model itself 
may be at fault. In particular in a model where the recombination front 
from doubly to singly ionised material hits the iron-rich inner core later, 
the secondary maxima would be dimmer and in better agreement with observations. 
However, investigating more modern explosion models is beyond the scope of 
this work and will be addressed in future studies.

\subsection{Comparison with other codes}
\label{sec:w7_codecomparison}
We now compare our bolometric and broad-band light curves with those obtained
with other radiative transfer codes for the W7 model. We start with the light
curves from the multi-energy group radiation hydrodynamics code \textsc{stella} 
\citep{Blinnikov2006} which treats the line opacity in the expansion opacity 
formalism \citep{Karp1977,Friend1983,Eastman1993}. E.~Sorokina (priv. comm.) 
provided us with bolometric, $U$, $B$, $V$, $R$ and $I$ light curves calculated 
with an updated version of this code using a line list of $\sim 1.6\cdot10^5$ 
lines. In this approach, matter is treated in LTE and redistribution is modelled
using an approximate source function. The light curves are shown in 
Figure~\ref{fig:w7lightcurves}: aside from having earlier peaks, even in 
bolometric light, these light curves are very similar to our light curves 
calculated in the simple ionisation treatment with atomic dataset cd23\_gf-5. 
This is not surprising since our simple ionisation description is appropriate 
for LTE and the atomic data sets are of comparable size. The obvious difference 
around the first peak in the $I$ band, which is dominated by the Ca\,{\sc ii} 
NIR triplet, is most likely due to our more complete treatment of fluorescence. 
Compared to our detailed ionisation treatment the same remarks as made in Section 
\ref{sec:w7_simplevsdetailed_ion} apply.

We also show light curves obtained with the 3D Monte Carlo radiative 
transfer code \textsc{sedona} \citep{Kasen2006}. \textsc{sedona} uses the expansion opacity
formalism but is capable of treating fluorescence with a more sophisticated 
approximation for a subset of lines using a downward-branching scheme. When
a photon is absorbed by a line treated in the expansion opacity formalism a 
two-level atom (TLA) approximation is used and the photon undergoes coherent 
scattering with probability $1-\epsilon$. Otherwise it is absorbed and 
reemitted at a wavelength sampled from the local thermal emissivity, thus 
representing both true absorption and fluorescence. 
In principle $\epsilon$ is a unique parameter for each line, however
\textsc{sedona} usually uses a common value for all lines [see \citet{Kasen2006} for a discussion].
Matter is treated in LTE with the local kinetic temperature 
being derived from balancing the thermal emissivity with the energy deposition 
by $\gamma$-ray heating and photon absorption recorded during the Monte Carlo 
simulations. The light curves shown in Figure~\ref{fig:w7lightcurves} were 
provided by D.~Kasen (priv. comm.) and use a line list of $\sim10^7$
lines treated in the TLA approximation.

The light curves are in good agreement with our calculations using the 
big\_gf-4 atomic dataset up to maximum light. After maximum light some 
differences manifest. The \textsc{sedona} $U$ and $B$ light curves after 50 days 
decline faster than ours -- this may be because of recombination to the 
neutral state, which is responsible for the relatively rapid decay of 
the $U$ and $B$ light curves obtained with our simple ionisation treatment (see 
Section~\ref{sec:w7_simplevsdetailed_ion}). Note that, despite using 
LTE conditions to describe the matter state, the \textsc{sedona} light curves 
do not fade as quickly after maximum light as those obtained with 
\textsc{stella} or with our simple ionisation treatment (which is equivalent to 
treating matter in LTE). This is likely to be due to the different manner 
in which the kinetic temperature is computed by the different codes.

The $V$ band light curves are almost identical. In the NIR light curves we get
remarkably similar results up to $\sim40$\,days. Afterwards, the \textsc{sedona}
light curves stay brighter. Since we have seen in Section \ref{sec:w7_simplevsdetailed_ion}
that a more complete recombination leads to an increased redistribution of
flux into the red and NIR by Fe\,{\sc i} and Co\,{\sc i} this is presumably 
due to a more complete recombination compared to our calculation. The strong 
differences in the $I$ band light curve are most likely due to the different 
treatment of line formation. The flux in the $I$ band is dominated by 
the Ca\,{\sc ii} NIR triplet for which the \textsc{sedona} light curves shown here 
assume pure scattering ($\epsilon=0$, see the discussion in \citealt{Kasen2006a}), 
while we treat the line formation in full detail.

\section{Application to an ellipsoidal toy model}
\label{sec:elmodel}
In this section we use an ellipsoidal toy model to demonstrate the 
multi-dimensional capabilities of the code and to illustrate the basic 
effects which large-scale ejecta asymmetries introduce in spectra and light 
curves. Large-scale asymmetries in the ejecta of SNe Ia are suggested both 
by observed polarimetry [see \citet{Wang2008} for a review] as well as 
by theoretical explosion models [e.g. from an off-centre ignition
condition \citep{Roepke2007b}, a deflagration to detonation transition 
\citep{Roepke2007} or in the gravitationally confined explosion scenario 
\citep{Plewa2004}]. We stress that the toy model we employ here is extremely
simple and has a much stronger asphericity than the observations or 
theoretical arguments suggest. However, it is a useful test case for our
code which clearly identifies the sense of the effects acting on the spectra
and light curves.

\subsection{The model}
Taking the total mass and composition of the W7 model and assuming homologous
expansion, we constructed a simple ellipsoidal toy model assuming rotational 
symmetry about the $z$-axis. The maximum velocities along the $x$- and $y$-axes 
are set to be only half the maximum velocity ($v_z^\mathrm{max}=27\,500\,\mathrm{km\,s}^{-1}$) 
along the $z$-axis, thus giving an axis-ratio of 1:1:2. Within the model, 
ellipsoidal surfaces are taken as surfaces of constant density and we adopt
a density profile 
  \begin{equation}
    \rho(v) \propto \exp\left(\sqrt{v_\mathrm{r}^2+\left(\frac{v_\mathrm{z}}{2}\right)^2}/v_0\right)
  \end{equation}
in cylindrical polar coordinates and $v_0=2750\,\mathrm{km\,s}^{-1}$. We assume a 
stratified composition with three zones. The innermost zone contains all 
the iron group material (Sc to Zn) and is surrounded by a zone of 
intermediate mass elements (F to Ca). The outermost zone contains the 
unburned material (He,C,N,O). Inside each zone the relative abundances 
are homogeneous and kept fixed at their W7 values.

We mapped this model to a $64^3$ grid and followed the propagation of 
$3.2\cdot10^7$ energy packets over 100 time steps from 2 to 80 days after the 
explosion, using the cd23\_gf-5 atomic dataset and the detailed ionisation treatment. 
The first 10 time steps have been calculated in the simple ionisation treatment
to get reliable initial values (see the discussion in Section \ref{sec:excandion}).
To save computational time we make use of our initial grey approximation setting $\tau_\mathrm{Th,min}=15 $ and $N_\mathrm{grey}=40$ (see Section \ref{sec:rpackets}).
Spectra and light curves were obtained by binning the escaping packets in 10
equal solid angle bins centred around the $z$-axis.

\subsection{Spectral evolution}
\begin{figure}
  \centering
  \includegraphics[angle=90]{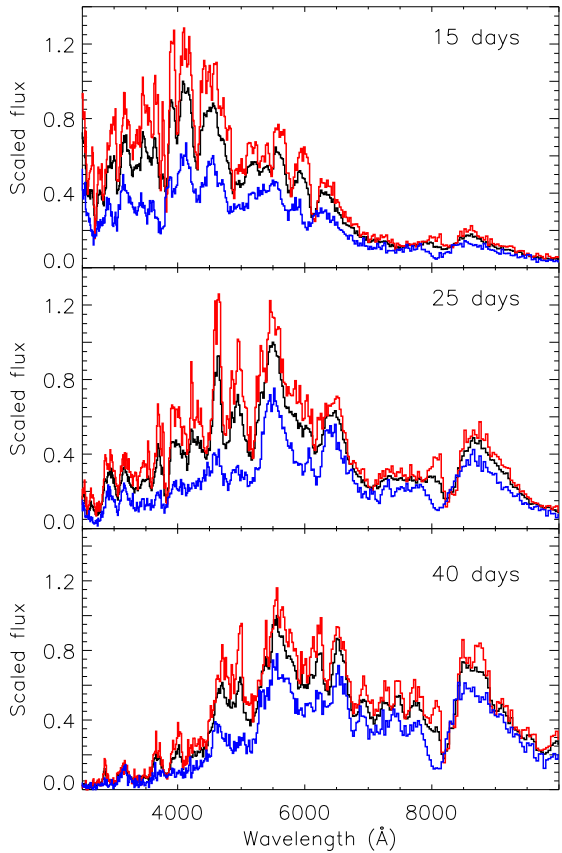}
  \caption{Spectra for the ellipsoidal toy model at 15, 25 and 40 days after 
    the explosion (from top to bottom). Blue/red lines are for viewing down 
    the major/minor axis. The black line shows an angle-averaged spectrum for 
    comparison.}
  \label{fig:elspectra}
\end{figure}

In Figure~\ref{fig:elspectra} we compare the 2500 to 10\,000 $\mathrm{\AA}$ 
spectra along the major and minor axes with an angle averaged spectrum for 
15 (around maximum light in B band), 25 and 40 days after explosion. In the 
absorption trough of the P Cyg feature of the Ca\,{\sc ii} NIR triplet one 
clearly sees the different velocity extent along the different axes: viewed 
down the major axis, the escaping photons see a velocity field twice as large
compared to the minor axes such that the absorption troughs extend farther 
into the blue. In general the spectra along the minor axes show much sharper
features than those along the major axis, where the higher velocities cause
stronger blending.

Furthermore the total flux observed along the minor axis is larger. This 
can be understood by a simple geometrical argument. Viewed down the minor 
axis the cross-section surface of our ellipsoid is twice as big as seen 
down the major axis. For an opaque ellipsoid of uniform surface brightness,
we would expect the same ratio for the flux (i.e. $\Delta M_\mathrm{geom} 
\approx -0.75$ in magnitudes). However, additional effects come into play 
such that the flux difference between major and minor axis depends on 
wavelength and time as discussed below.

\subsection{Broad-band light curves}
\begin{table}
  \centering
  \caption{$\Delta M=M_\mathrm{minor}-M_\mathrm{major}$ for selected bands in the ellipsoidal
    model at different times after explosion.}
  \label{tab:el_deltam}
  \begin{tabular}{@{}lcccccc@{}}
  \hline
    $t$           & 15d   & 22d   & 29d   & 35d   & 42d   & 49d\\
  \hline
    $\Delta M(U)$ & -0.93 & -1.01 & -0.92 & -0.80 & -0.70 & -0.87\\
    $\Delta M(V)$ & -0.52 & -0.65 & -0.54 & -0.48 & -0.45 & -0.46\\
    $\Delta M(I)$ & -0.42 & -0.41 & -0.35 & -0.36 & -0.34 & -0.23\\
    $\Delta M(H)$ & -0.38 & -0.24 & -0.22 & -0.27 & -0.27 & -0.19\\
  \hline
  \end{tabular}
\end{table}

\begin{figure*}
  \centering
  \includegraphics[angle=90]{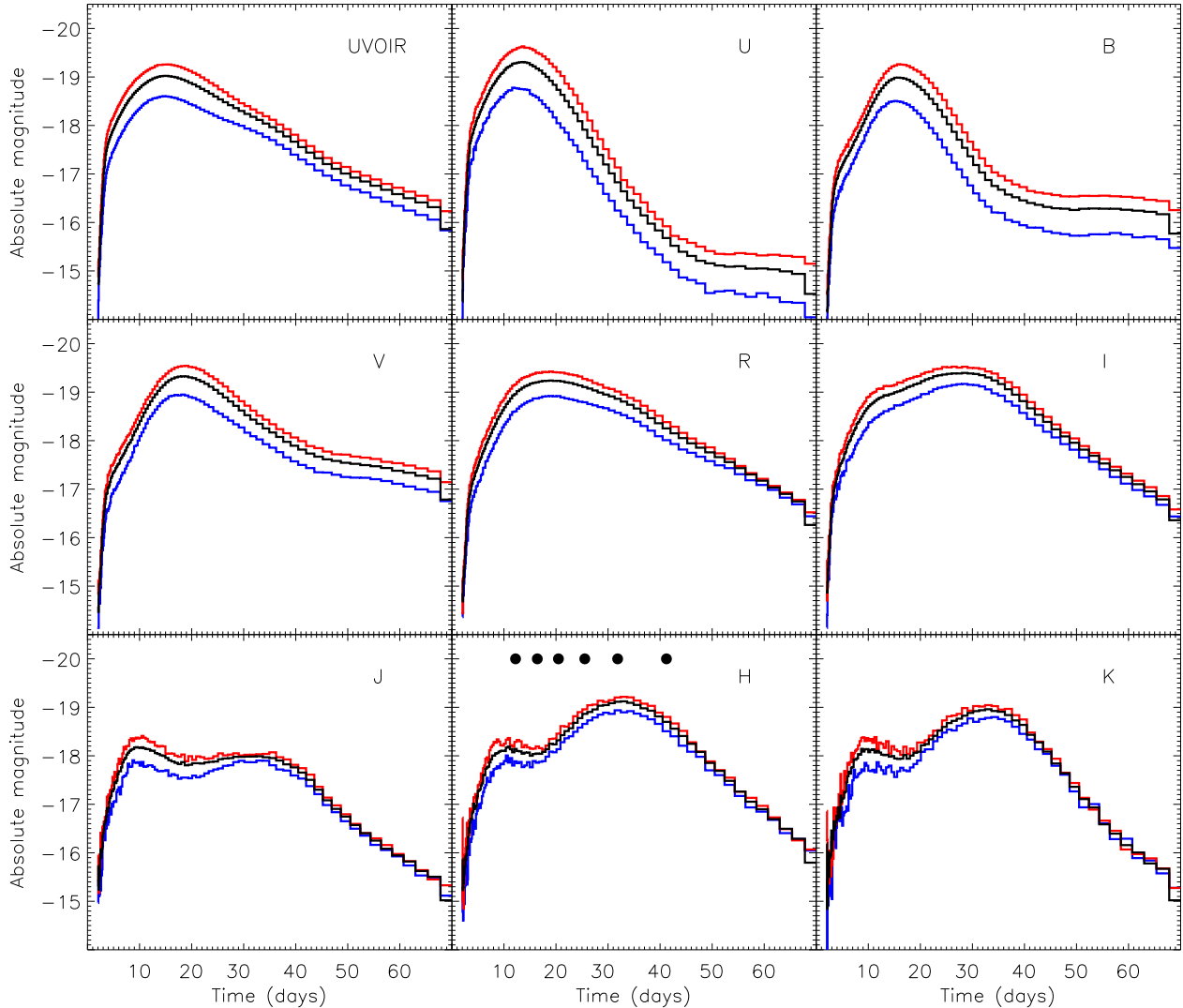}
  \caption{\textit{UVOIR} bolometric and \textit{U,B,V,R,I,J,H,K} light curves
    for the ellipsoidal toy model. The blue/red lines are the light curves 
    obtained along the major/minor axis. The black line shows an angle-averaged
    light curve for comparison. The filled circles in the $H$-band panel indicate
    the times for which snapshots of the ionisation state of Fe are shown in
    Figure~\ref{fig:el_ionstruct}.}
  \label{fig:el_lc}
\end{figure*}

The broad-band light curves are shown in Figure~\ref{fig:el_lc}. As
expected, the light curves observed along the minor axes are always
brighter than those observed along the major axis. The viewing-angle
effect is always strongest in the bluer bands and after maximum light 
becomes weaker with time in $V$ and redder bands (see Table~\ref{tab:el_deltam})
as the ejecta become optically thin in these parts of the spectrum.

\begin{figure*}
  \centering
  \includegraphics[angle=90]{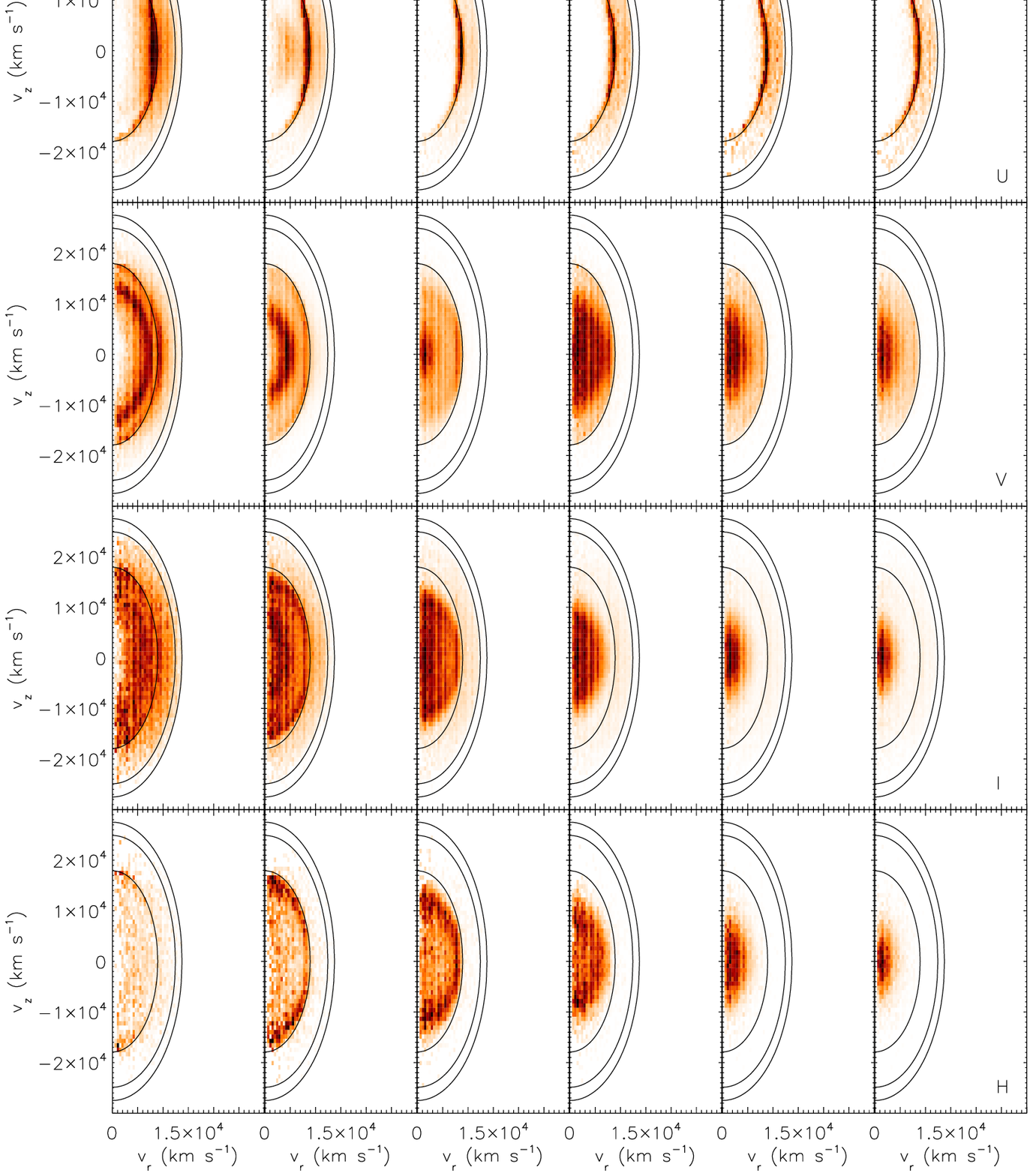}
  \caption{Region of last emission for selected bands ($U$, $V$, $I$, $H$ 
    from top to bottom) and different times (from left to right). Dark 
    regions contribute most to the flux escaping in the band. The solid 
    lines indicate the three zones of different composition, with Sc--Zn 
    in the centre surrounded by F--Ca and finally He,C,N and O in the 
    outer zone. The model is symmetric under rotation about the $z$-axis.}
  \label{fig:el_em}
\end{figure*}

In the $U$ and $B$ bands, which stay optically thick throughout our
calculation, this effect remains significant until the latest times 
of our simulation and the difference $\Delta M=M_\mathrm{minor}-M_\mathrm{major}$
is even bigger than one would expect from the simple geometrical 
argument (which would imply $\Delta M_\mathrm{geom} \approx -0.75$; 
see above). This is simply because we do not have an opaque ellipsoid of 
uniform surface brightness. In fact, Figure~\ref{fig:el_em} -- which 
illustrates the region of last emission (RLE) of selected bands at 
different times -- shows that the $U$ band RLE is concentrated around 
the equatorial plane which acts to amplify the geometrical effect (as 
long as the ejecta are optically thick, the direction of escape tends 
to be peaked normal to the contours of constant density). This concentration
is a consequence of the large number of highly optically thick Fe group 
lines in the UV: $U$ band photons are more strongly trapped than photons 
in other bands and therefore tend to preferentially leak out along the 
equatorial plane where the velocity swept out is smallest.

\begin{table}
  \centering
  \caption{Peak times and $\Delta M_{15}$ for selected bands in the ellipsoidal model.}
  \label{tab:el_peakdm15}
  \begin{tabular}{@{}lcccccc@{}}
  \hline
        & $t_\mathrm{max}^\mathrm{major}$ (d) & $t_\mathrm{max}^\mathrm{minor}$ (d) 
        & $\Delta M_{15}^\mathrm{major}$         & $\Delta M_{15}^\mathrm{minor}$  \\
  \hline
    $UVOIR$ & 14.7 & 15.8 & -0.6 & -0.9\\
    $U$     & 11.8 & 13.7 & -1.7 & -1.9\\
    $B$     & 15.2 & 15.8 & -1.7 & -1.8\\
    $V$     & 18.3 & 19.0 & -0.9 & -1.1\\
    $R$     & 19.0 & 19.0 & -0.5 & -0.6\\
  \hline
  \end{tabular}
\end{table}

In contrast, the RLE of $V$ and redder bands do not show a strong enhancement 
around the equatorial plane. The $V$ band RLE around maximum light is nearly 
ellipsoidal, making $\Delta M$ close to $\Delta M_\mathrm{geom}$. At later 
times, the emission becomes more isotropic as the ejecta become optically 
thinner (in Figure~\ref{fig:el_em} the whole ellipsoid contributes to the RLE) 
and $\Delta M$ decreases. For the $I$ and $H$ band, in which $\Delta M$ is 
always less than $\Delta M_\mathrm{geom}$, the ejecta are already becoming
optically thin around maximum light. 

The $H$ band RLE shows a slight enhancement towards the polar regions 
around maximum light. This is related to the reduced $U$ band emission in these 
regions. The high optical depths for blue and UV photons mean that 
fluorescence redistributes flux into the NIR where optical depths are 
lower such that the radiation can escape.

In general the NIR light curves are least viewing-angle dependent, thus 
supporting the use of NIR light curves for cosmological distance 
measurements since less intrinsic scatter would be expected if 
geometry effects have any role in the observed properties of SNe 
Ia. However more detailed studies are needed to investigate this.

The light curves observed along the major axis peak slightly earlier than
those observed along the minor axes (see Table~\ref{tab:el_peakdm15}). 
Furthermore light curves observed along the major axis decline more slowly 
than those observed along the minor axes (compare the 
$\Delta M_{15}$\footnote{$\Delta M_{15}$ gives the change in
magnitude between maximum light and 15 days after maximum light.} values 
in Table~\ref{tab:el_peakdm15}). A similar effect was already found in 
a study using a grey version of our code by \citet{Sim2007}, and we 
note that it is opposite to the sense of the observed light curve 
width-luminosity relation.

The plot showing the regions of last emission in Figure~\ref{fig:el_em} also
traces the stratified composition of our model. The $U$ and $I$ bands -- which
have non-negligible contributions of Ca due to the Ca\,{\sc ii} H and K lines 
in $U$ and the Ca\,{\sc ii} NIR triplet in $I$ -- show significant emission 
from the zone of intermediate mass elements. In contrast, the RLE in the $V$ 
and $H$ bands -- which are dominated by Fe group elements -- is concentrated 
towards the iron-rich inner core.

\subsection{Secondary maximum in the NIR bands}

\begin{figure*}
  \centering
  \includegraphics[angle=90]{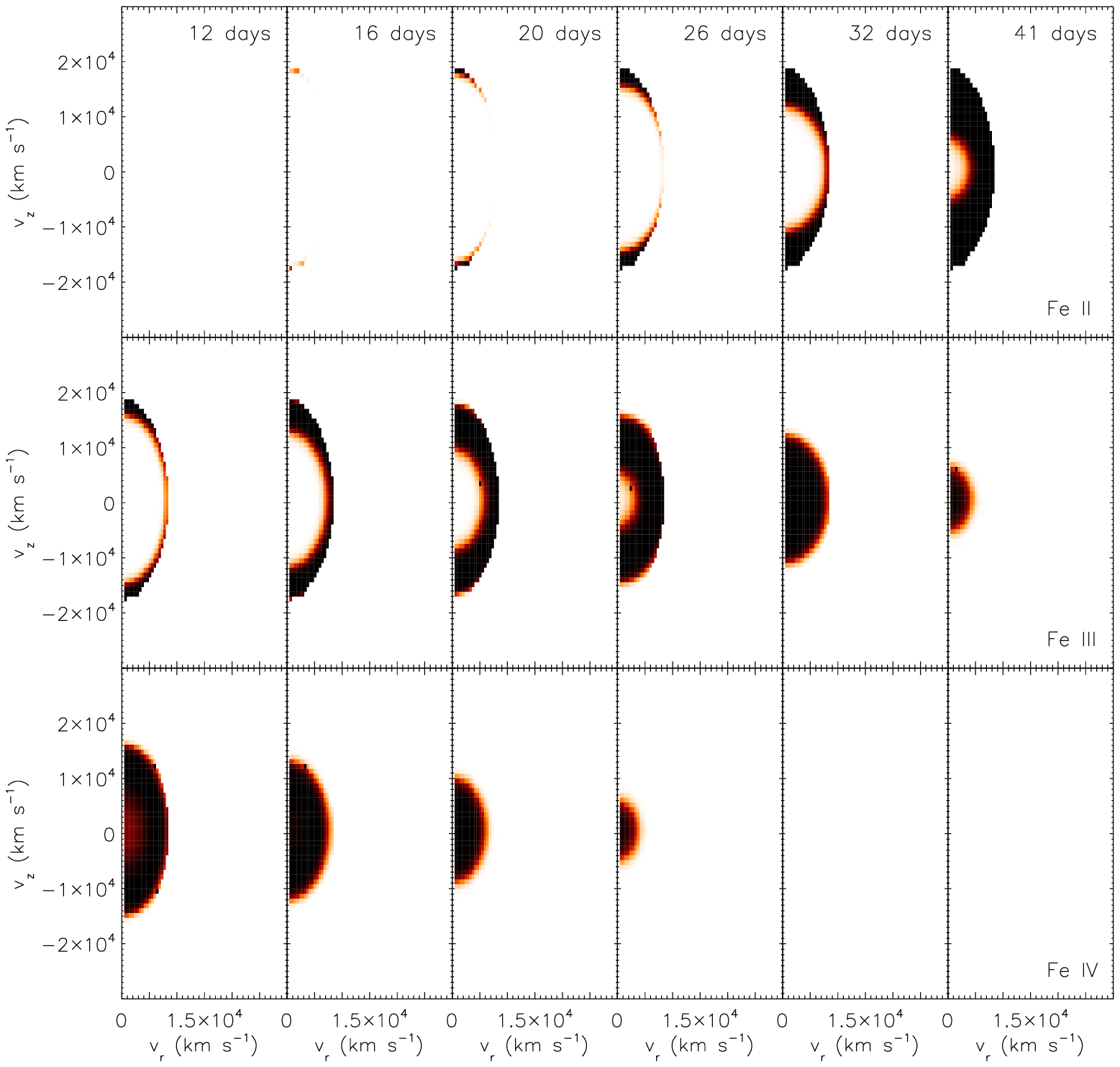}
  \caption{Ionisation fractions of Fe\,{\sc ii}, Fe\,{\sc iii} and Fe\,{\sc iv} 
    (from top to bottom) for selected times (from left to right). The corresponding
    times are marked with filled circles in the $H$ band panel of Figure~\ref{fig:el_lc}.
    Black regions have an ionisation fraction of 1. Lower ionisation fractions are 
    shown in lighter reddish colours down to white (0) on a linear scale.}
  \label{fig:el_ionstruct}
\end{figure*}

As discussed by \citet{Kasen2006a}, the secondary maximum in the NIR 
bands forms when the inner iron-rich core recombines from doubly ionised 
iron group elements to the singly ionised state. With the singly ionised 
state becoming dominant, the number of optically thick lines in the blue 
increases, thus blocking blue radiation very effectively and making the 
redistribution of flux from the UV and blue part of the spectrum into 
the red and NIR by fluorescence very effective. Here we discuss this 
effect for the $H$ band light curve and show the ionisation fractions of 
Fe\,{\sc ii--iv} (which we use as a proxy for all iron group elements) in 
Figure~\ref{fig:el_ionstruct} for selected times from near the first peak
($\sim$ 10 days after explosion) to well after the second peak in the $H$ 
band ($\sim$ 32 days after explosion, compare Figure~\ref{fig:el_lc}).

Around the first peak the iron-rich core consists predominantly of 
Fe\,{\sc iv}. In the very centre there is still a contribution of 
Fe\,{\sc v} (not shown in Figure~\ref{fig:el_ionstruct}) and
at the outer edge there is a small region which has already recombined 
to Fe\,{\sc iii}. As time passes the ejecta cool and the Fe\,{\sc iv}
region recedes. At the same time the Fe\,{\sc iii} region extends further 
in and a contribution of Fe\,{\sc ii} starts to build up around the
outsides. Recombination occurs first at the poles since they see the 
least amount of ionising radiation (compare Figure \ref{fig:el_em}). 
The secondary maximum in the $H$ band occurs at $\sim 32$ days when 
the Fe\,{\sc ii} region first forms a complete elliptical ring 
(Figure~\ref{fig:el_ionstruct}). This is the point at which the opportunity 
for fluorescence is maximal. Note that although the ionisation fractions
vary with position, the NIR bands are sufficiently optically thin 
that the NIR flux is approximately angle-independent at these epochs
(this is apparent from Figure~\ref{fig:el_em} where the full ellipsoid
is bright). At late times the recombination front moves inwards as the 
light curves fade until only a small Fe\,{\sc iii} core remains at the 
end of our simulation.

Comparing the two different lines-of-sight in Figure~\ref{fig:el_lc}, 
we see that the local minimum between the first and secondary
maximum in the $H$ band light curve along the polar axis occurs
slightly earlier ($\sim 15$ days) than along the equatorial axes 
($\sim 18$ days). This  follows from the earlier recombination 
from Fe\,{\sc iii} to Fe\,{\sc ii} in polar lobes which gives an 
earlier increase in the redistribution of flux from the UV and blue 
into the red and NIR.

\section{Conclusions}
\label{sec:conclusions}
Based on the approach of \citet{Lucy2002,Lucy2003} we extended the grey 
time-dependent 3D Monte Carlo radiative transfer code of \citet{Sim2007} 
to a non-grey opacity treatment. The new code, \textsc{artis}, treats the 
$\gamma$-deposition and spectrum formation in detail and solves the ionisation 
balance together with the thermal balance equation consistently with the 
radiation field. Line formation is treated in the Sobolev approximation using 
the macro-atom approach of \citet{Lucy2002,Lucy2003} to model atomic physics in 
detail. This allows parameter-free radiative transfer simulations for 3D hydro 
models with a maximum of predictive power to be made.

We applied this code to the well-studied one-dimensional deflagration model 
W7 \citep{Nomoto1984} as a test case and found good agreement with both 
earlier work using different codes [\textsc{sedona} \citep{Kasen2006}; \textsc{stella} 
\citep{Blinnikov2002,Blinnikov2006}] and observations [SN 1994D 
\citep{Patat1996}; SN 2001el \citep{Krisciunas2003}]. Concerning the NIR 
light curves, we confirm the importance of line fluorescence in modelling 
these bands and, particularly their strong dependence on the size of the 
atomic data set in use. Rather than adopting LTE, we assume photoionisation 
equilibrium and solve the ionisation balance equations using rates 
consistent with the radiation field. We showed that this leads to 
significant differences in the ionisation structure after maximum light, 
which strongly affect spectra and light curves. The macro-atom formalism 
allows us to avoid introducing a parameterised treatment of line fluorescence, 
such that we account for the differing contributions of resonance scattering 
and fluorescence in different lines (see Figure~\ref{fig:redmat_lines}). 

To demonstrate the multi-dimensional capabilities of the code we have presented
calculations for an ellipsoidal toy model as an example. As expected [e.g. from 
the grey study by \citet{Sim2007} for a similar model], light curves observed
along the minor axes are brighter than those observed along the major axis.
The sensitivity decreases with time, as the ejecta become less optically
thick, and from blue to red wavelengths. If line-of-sight effects due to 
asymmetric explosions have a significant influence on the scatter around 
the Phillips relation of SNe Ia \citep{Phillips1993}, this could explain 
why the NIR light curves seem to be more homogeneous and thus most promising
for cosmological distance measurements. However more detailed studies are 
needed to investigate this fully for realistic models.

In future work we will extend our method to include a treatment of non-thermal
excitation and ionisation which we currently do not treat in detail but which 
can be important \citep[for a discussion see][]{Baron1996a}. This will also 
allow us to  extend the simulations to  later epochs (nebular phase) which are 
strongly affected by these processes. Eventually time-dependent terms in the 
statistical equilibrium equations could be taken into account and polarisation 
could be treated, thereby extending the range of data with which we can make 
meaningful comparisons.

\section*{Acknowledgments}
We are very grateful to W.~Hillebrandt for stimulating discussions and 
making this work possible. Furthermore we thank E.~Sorokina, S.~Blinnikov 
and D.~Kasen for providing us with results of light curve calculations from
\textsc{stella} and \textsc{sedona} and also for useful comments. We also acknowledge helpful 
discussions with S.~Hachinger, F.~R\"opke, D.~Sauer and S.~Taubenberger.
We thank the anonymous referee for several helpful suggestions.\\
This work made use of the SUSPECT database.

\bibliographystyle{mn2e}   
\bibliography{literature}

\bsp

\label{lastpage}

\end{document}